\documentclass[prb,twocolumn,amsmath,amssymb,superscriptaddress]{revtex4}
\usepackage[T1]{fontenc}
\usepackage[latin1]{inputenc}
\usepackage{graphics}
\usepackage{amssymb}
\usepackage{amsmath}
\usepackage{overpic}

\newcommand{\be}{\begin{equation}}
\newcommand{\ee}{\end{equation}}
\newcommand{\bea}{\begin{eqnarray}}
\newcommand{\eea}{\end{eqnarray}}

\def\a{\alpha}
\def\b{\beta}
\def\e{\varepsilon}
\def\d{\delta}
\def\g{\gamma}
\def\m{\mu}
\def\l{\lambda}

\def\n{\nu}
\def\o{\omega}
\def\p{\pi}
\def\s{\sigma}

\def\G{\Gamma}
\def\D{\Delta}
\def\O{\Omega}
\def\L{\Lambda}

\def\ra{\rightarrow}
\def\up{\uparrow}

\def\down{\downarrow}

\def\Ra{\Rightarrow}
\def\pd{\partial}

\def\bk{{\bf k}}

\def\bq{{\bf q}}
\def\bs{{\bf s}}

\def\bu{{\bf u}}
\def\bV{{\bf V}}
\def\bx{{\bf x}}

\def\DD{{\cal{D}}}

\def\nn{\nonumber}
\def\lb{\label}
\def\pref#1{(\ref{#1})}

\newcount\bozza \bozza=0
\ifnum\bozza=1
\newdimen\shift \shift=-2truecm
\def\lb#1{%
{\label{#1}\rlap{\kern\shift{$\scriptstyle#1$}}}}
\else\def\lb#1{\label{#1}} \fi

\begin{document}

\title{Leggett modes in iron-based superconductors as a probe of \\Time Reversal
Symmetry Breaking}

\author{M. Marciani}
\affiliation
{Institute for Complex Systems (ISC), CNR, U.O.S. Sapienza and \\
Department of Physics, Sapienza University of Rome, P.le A. Moro 2,
00185 Rome, Italy}

\author{L. Fanfarillo}
\affiliation{Instituto de Ciencia de Materiales de Madrid, ICMM-CSIC,
  \\
Cantoblanco, E-28049 Madrid, Spain} 
\affiliation
{Institute for Complex Systems (ISC), CNR, U.O.S. Sapienza and \\
Department of Physics, Sapienza University of Rome, P.le A. Moro 2,
00185 Rome, Italy}

\author{C. Castellani}
\affiliation
{Institute for Complex Systems (ISC), CNR, U.O.S. Sapienza and \\
Department of Physics, Sapienza University of Rome, P.le A. Moro 2,
00185 Rome, Italy}

\author{L. Benfatto}
\affiliation
{Institute for Complex Systems (ISC), CNR, U.O.S. Sapienza and \\
Department of Physics, Sapienza University of Rome, P.le A. Moro 2,
00185 Rome, Italy}

\date{\today}

\begin{abstract}
  Since their discovery, it has been suggested that pairing in
  pnictides can be mediated by spin fluctuations between hole and electron
  bands.  In this view, multiband superconductivity would
  substantially differ from other systems like MgB$_2$, where pairing
  is predominantly intraband. Indeed, interband-dominated pairing
  leads to the coexistence of bonding and antibonding superconducting
  channels. Here we show that this has profound consequences on the nature
  of the low-energy superconducting collective modes. In particular,
  the so-called Leggett mode for phase fluctuations is absent in the
  usual two-band description of pnictides. On the other hand, when
  also the repulsion between the hole bands is taken into account, a
  more general three-band description should be used, and a Leggett
  mode is then allowed. Such a model, that has been proposed for
  strongly hole-doped 122 compounds, can also admit a low-temperature
  $s+is$ phase which breaks the time reversal symmetry. We show
  that the (quantum and thermal) transition from the ordinary
  superconductor to the $s+is$ state is accompanied by the vanishing of the mass of
  Leggett-like phase fluctuations, regardless the specific values of
  the interaction parameters. This general result can be obtained by
  means of a generalized construction of the effective action for the
  collective degrees of freedom that allows us also to deal with the
  non-trivial case of dominant interband pairing.
\end{abstract}


\maketitle

\section{Introduction}

At microscopic level the appearance of superconductivity requires the
pairing of electrons into Cooper pairs, which can then form a macroscopic
coherent state responsible for the superfluid behavior.  Within BCS theory,
which successfully explained the superconducting (SC) phenomenon in the
so-called conventional superconductors, electrons can overcome their mutual
repulsion thanks to the presence of phonons, which overscreen the Coulomb
repulsion leading to a residual attraction responsible for the
pairing. However, such a mechanism poses an upper limit to the attainable
transition temperature, that has been widely exceeded in the so-called
high-temperature superconductors, as cuprate or iron-based
systems.\cite{mazin_nat10,basov_natphys11} In all
these cases superconductivity emerges and/or competes with strong
electron-electron repulsion, that can be accommodated by Cooper pairs by
means of an unconventional form of the wave function, as it is the case in
cuprates, where the $d$-wave symmetry of pairing allows the pairs to
overcome the on-site Hubbard-like repulsion. In the case of pnictides the
mechanism is somehow similar, once that the multiband nature of the Fermi
surface is taken into account. Indeed, it has been suggested by several
microscopic approaches\cite{hirschfeld_review11} that at low energy the intraband Coulomb repulsion
is overcome by the interband repulsion, which allows the pairs to be formed
in different bands with a gap having opposite sign, the so-called $s^\pm$
symmetry. Roughly speaking, such a sign change converts a repulsion in
attraction, making the pairs formation possible. Notice that such a pairing
mechanism is fundamentally different from the one observed in other
multiband superconductors, as e.g. MgB$_2$. Here indeed the largest pairing
channel is the intraband phononic one,\cite{review_mgb2} and the interband interaction is
only responsible for a relatively small Josephson-like coupling of pairs in
different bands. In this respect, pnictide superconductors represent a
completely different class of SC systems with respect to MgB$_2$.

A fundamental question associated to the unconventional nature of pairing
is how it can affect the behavior of the SC collective modes, which in turn
can influence the observable physical quantities, giving indirect
information on the nature of the underlying SC state. Such an issue has
been widely discussed in the past within the context of cuprate
superconductors,\cite{depalo_prb00,randeria_prb00,benfatto_prb01,dorsey_prl01,sharapov_prb01,sharapov_prb02,benfatto_prb04}
and it has been the subject of intense investigation in
the recent literature on pnictide
superconductors.\cite{burnell_prb10,ota_prb11,babaev_prb11,hu_prl12,stanev_prb12,maiti_prb13}
 Here the issue is made
even more involved by the presence of several bands, that would suggest the
presence of multiple collective modes associated to the fluctuations of the
amplitude and phase of the condensates in the various bands. For example,
it has been discussed\cite{burnell_prb10,hu_prl12} the possibility to observe the so-called Leggett mode,\cite{legget}
that corresponds to the relative density (phase) fluctuations of the
condensate in the various bands. As it has been shown long ago in a seminal
paper by Leggett,\cite{legget} such a massive mode could eventually lie below the
threshold for particle-hole excitations, avoiding then its overdamping. Such
a situation is partly realized in MgB$_2$,\cite{sharapov_epjb02,efetov_prb07,blumberg_prl07,klein_prb10} where indeed experimental
signatures of the Leggett mode have been identified in Raman
spectroscopy.\cite{blumberg_prl07} 
In the case of pnictides it has been also suggested the intriguing
possibility that the Leggett
mode\cite{babaev_prb11,hu_prl12,maiti_prb13} 
becomes massless at the quantum transition between
an ordinary $s^\pm$ state and a time-reversal-symmetry broken (TRSB) state.
Such a TRSB state can emerge for example in a three-band case when
interband repulsion is equally large between all the
bands:\cite{stanev_prb10,thomale_prl11,suzuki_prb11,chubukov_prb12} in this
situation the sign change between one band and the remaining two is
frustrated, leading to an intrinsically complex order parameter
$(\D_1^*,\D_2^*,\D_3^*)\neq (\D_1,\D_2,\D_3)$.\cite{stanev_prb10} Since the emergence of a massless
collective mode could bear several observable consequences in physical
observable, as e.g. Raman response\cite{hu_prl12} or intervortex
interactions,\cite{babaev_prb11} it could be used as a smoking gun to test
the appearance or not of a TRSB state in pnictides.

Quite interestingly, the theoretical investigation of the properties of
collective modes in pnictides suffered until now of a fundamental
limitation. Indeed, as we discussed at the beginning, pairing in pnictides
arise mainly from {\em interband} interactions. However, very often a
modelization has been used in the literature based on multiband models with
predominant {\em intraband} pairing
interactions.\cite{ota_prb11,babaev_prb11,hu_prl12}  While this makes it
possible to derive the collective modes using standard procedures based on
the construction of the effective action for the collective degrees of
freedom,\cite{ota_prb11,hu_prl12} it makes these results unsuitable for the specific case of
pnictides. On the other hand, an alternative derivation based on the direct
diagrammatic derivation of the collective response functions, as the one
used in Refs.\ \cite{stanev_prb12,maiti_prb13}, does not allow for a simple
general understanding of the number and nature of the collective modes.  As
we discuss in the present paper, the difference between the two cases is
not only quantitative but qualitative. Indeed, when interband interaction
dominates, as it is the case physically relevant for pnictides, the number
itself of available low-energy collective modes is smaller than the number of bands
involved in the problem. In this case the correct understanding of
the SC collective modes should be based on the number of {\em SC bonding channels},
that is usually smaller than the number of bands involved. This
leads to several profound differences between pnictides and ordinary
(intraband-dominated) multiband superconductors, like e.g. MgB$_2$.

A powerful root to enlighten these differences is the explicit
construction of the action for the collective modes starting from a
microscopic model for pnictides that assumes predominant interband
pairing. In the ordinary case of intraband-dominated pairing such a
procedure relies on the use of the so-called Hubbard-Stratonovich (HS)
decoupling of the SC interaction by means of a bosonic fields associated
to the pairing operators.\cite{nota_HS} This approach has been
successfully applied to two\cite{sharapov_epjb02}  or three-band
\cite{ota_prb11,hu_prl12} models 
with predominant intraband pairing. However, when interband coupling
dominates, as it is the case for pnictides, the HS decoupling must be properly modified to account for
the presence of antibonding SC channels, an issue
that has been often overlooked in the recent
literature in the derivation of effective functionals
both above\cite{maiti_prb13,fernandes_cm13} and below \cite{burnell_prb10}
$T_c$.  Here we follow instead the strategy outlined recently in Ref.\
[\onlinecite{fanfarillo_prb09}], where the correct implementation of the HS
procedure has been used to describe the fluctuations above $T_c$. 
We then introduce a transformation of the pairing fields in the various
bands that allows us to show that below $T_c$ the fluctuations 
associated to the antibonding SC channels do not give rise to observable
collective modes. This result follows immediately from a general
correspondence between the
low-energy collective phase fluctuations and the multiband mean-field
equations. When applied
to the two- or three-band case with dominant interband pairing,
relevant for pnictides,  this
correspondence allows one to show that: (i) in the two-band case the
Leggett mode is absent, in contrast to intraband-dominated superconductors
as MgB$_2$;  (ii) in the three-band case a Leggett mode is
present, it becomes massless at the TRSB transition and it acquires
again a small mass inside the TRSB phase due to the mixing to
amplitude fluctuations. A second low-energy mode appears in the TRSB
state, even though it is usually found very near to the threshold for
single-particle excitations.  In contrast to the previous literature,
which focused on the softening of the Leggett mode at $T=0$ as a
function of the SC coupling leading the system through a quantum phase
transition to a TRSB state,\cite{hu_prl12,stanev_prb12,maiti_prb13} we
discuss its occurrence as a function of temperature. Indeed, the
thermal phase transition between a TRS and TRSB phase is possibly
realized in a much wider range of parameters for realistic systems,
and then it has definitively more chances to be observed
experimentally.

The structure of the paper is the following. In Sec. II-A we outline
the main steps that lead to the effective action for the collective
degrees of freedom starting from a microscopic two-band model with
interband-dominated pairing. The character of the amplitude and phase
modes is discussed in Sec. II-B, where we also show the absence of the
ordinary Leggett mode, found instead in two-band superconductors with
intraband-dominated pairing. The three-band case is discussed in
Sec. III. Sec. III-A is devoted to a brief review of the possible
relevance for pnictides of three-band models which admit a TRSB
state. The general structure of the collective modes is discussed in
Sec. III-B, where it is established the correspondence between the
TRSB transition and the vanishing of the mass of a Leggett-like
mode. In Sec. III-C we consider a specific set of SC couplings to show
explicitly the temperature (and quantum) evolution of the low-energy
modes across the TRSB transition. The results of Fig.\ \ref{fig-l112}
summarize the main physical messages relevant for the reader who is
not interested in the theoretical aspects of their derivation, and
Sec. III-D contains a general discussion on the experimental probes
that can be used to test the behavior of the phase collective modes
near the TRSB state. 
Sec. IV contains our final remarks and the summary of the main results of the
paper. Additional technical details, which are useful to make a
direct comparison with previous work in the literature, are reported
in the appendices. Appendix A shows the equivalence between the
derivation of the Gaussian action for SC fluctuations 
done in polar or cartesian coordinates. Appendix B discusses the
two-band case with dominant intraband pairing by means of the
formalism of the present manuscript. Finally, in Appendix C we discuss
the general connection between the TRSB transition and the
effective action for the three-band model.

\section{The effective action for a two-band model}
\label{2bands}

\subsection{Construction of the effective action}
To show explicitly the peculiar role of interband interactions in
determining the nature of the collective modes we first describe the
two-band case. Having in mind pnictide systems, such an effective
modelization is usually appropriate for  systems not too far
away from half-filling.  Indeed, in this case one can assume that the most
relevant interactions are between the two hole pockets centered at $\Gamma$
and the two electron ones centered at M, with no interaction between the
hole bands (see also discussion in Sec. IIIA below).  Assuming also that
the electron bands are degenerate, this four-band model can be
mapped\cite{benfatto_prb08,fanfarillo_prb09} into an effective
BCS-like two-band one as
\begin{eqnarray}
H&=&H_0+H_{int},\nn\\
H_{0} & = & \sum_{\bk,l,\sigma} \xi^l_\bk c^{l\dagger}_{\bk\s}c^l_{\bk\s}\nn\\
\lb{ham2}
H_{int} & = & -\sum_{\bq,ij} \hat g_{lm} \phi_{l,\bq}^{\dagger}\phi_{m,\bq}\qquad(l,m=1,\,2)
\end{eqnarray}
where 
\be
\lb{defffields}
\phi_{l,\bq}= \sum_\bk c^l_{\bk+q\downarrow}c^l_{-\bk\uparrow}
\ee
is the pairing operator in each band and the
matrix $\hat g_{lm}$
\be
\lb{defg2}
\hat g =\begin{pmatrix}\alpha & \gamma\\
\gamma & \beta
\end{pmatrix}, \quad \mathrm{det}\, \hat g<0
\ee
describes predominant interband pairing. The bare
electronic dispersion in Eq.\ \pref{ham2} will be approximated with a
parabolic one, $\xi^l_\bk=\e^l_0\pm\bk^2/2m_l-\mu$, with the plus or minus
sign for electrons or holes, respectively, and the chemical potential $\mu$
will be taken equal to zero. We notice that while to account quantitatively
for the correct spectral and thermodynamic properties of pnictides a more
refined Eliashberg-like multiband approach is
needed,\cite{benfatto_prb09} the Hamiltonian \pref{ham2} can be considered
an appropriate starting point to discuss the general structure of
collective modes in most pnictides.

As customary, the microscopic effective model for 
the collective modes can be derived by considering the action
corresponding to the Hamiltonian \pref{ham2}, within the 
finite-temperature Matsubara formalism, 
\be  
S=\int^{\beta}_{0} d\tau\left\{\sum_{l,\mathbf{k}\sigma}
c ^{l\dagger}_{\mathbf{k}\sigma}(\tau)[\partial_{\tau}+\xi_{\bk}]
c ^l_{\mathbf{k}\sigma}(\tau)d\tau+H_I(\tau)\right\}, \label{smicro} 
\ee 
where $\tau$ is the imaginary time and $\beta=1/T$.
To obtain the effective action in terms of the order-parameter 
collective degrees of freedom, the interaction $H_I$ is usually decoupled in the 
particle-particle channel by means of the Hubbard-Stratonovich\cite{hubbard_prl59} field $h_{HS}$:
\bea
\lb{hs}
e^{\pm\Lambda \phi^\dagger\phi}&=&\int \DD h_{HS}\, 
e^{-|h_{HS}|^2/\Lambda+
\sqrt{\pm 1}(\phi^\dagger h_{HS}+h.c.)}.
\eea
In the above equation the imaginary unit $\sqrt{-1}\equiv i$ signals the presence of a
repulsive particle-particle interaction. In the usual single-band case\cite{nota_HS}
one deals with an interaction attractive in the particle-particle
channel, so no imaginary unit appears. However in the present multiband
case with predominant interband coupling the diagonalization of the $\hat g$
matrix with a proper rotation $R$ will lead in general also to a {\em
  negative} eigenvalue, corresponding to repulsion in the
particle-particle channel:
\be
\lb{gdiag}
\hat g= R^{-1}\hat\Lambda R=
R^{-1}
\begin{pmatrix}\Lambda_{1} & 0\\
0 & -\Lambda_{2}
\end{pmatrix}R, \quad \L_{1,2}>0.
\ee
As we shall see below, the saddle-point values of the HS fields $h_{HS}$ are
connected to the SC gaps in the various bands. However, the imaginary unit
in the transformation \pref{hs} would force us to shift the integration
contour of $Re h_2$ by a finite imaginary quantity, so that $Re h_2
\,\in \, \mathbb{R}+iA$, see discussion below Eq.\ \pref{sfl}. 
To preserve an ordinary integration contour we will enforce $A=0$  by
taking advantage of the fact that the interaction Hamiltonian
$H_I$ can be put in the diagonal form under a more
general transformation $T=\hat H_\varphi R$ (with $\mathrm{det} \, T=1$), where the matrix $\hat H_\varphi$
\begin{equation}
\lb{defhp}
\hat H_{\varphi}=\begin{pmatrix}
1/\sqrt{\Lambda_{1}} & 0\\
0 & 1/\sqrt{\Lambda_{2}}
\end{pmatrix}\begin{pmatrix}\cosh\varphi & \sinh\varphi\\
\sinh\varphi & \cosh\varphi
\end{pmatrix}\begin{pmatrix}\sqrt{\Lambda_{1}} & 0\\
0 & \sqrt{\Lambda_{2}}
\end{pmatrix}
\end{equation}
leaves $\hat \L$ invariant:
\be
\lb{deft}
\quad  
\hat H_\varphi^T\hat \L \hat H_\varphi =\hat \L.
\ee
As one can see, the $\hat H_\varphi$ matrix is essentially proportional to
the matrix of hyperbolic rotations, which commutes with the $diag(1,-1)$
matrix which arises when the two eigenvalues of $\hat g$ have opposite
sign. The relation \pref{deft} holds regardless the value of the parameter $\varphi$,
which will be chosen to decouple the two SC channels, see Eq.\ \pref{fix}
below. Indeed, thanks to Eq.\ \pref{deft} $\hat g$ can
be diagonalized by $T$ as well:
\be
\lb{grot}
T=\hat H_\varphi R \quad \Ra \quad \hat g= T^T \hat \L  T, \quad \hat\L^{-1}=T\hat g^{-1}T^T.
\ee
Thus, if we introduce the new combinations of fermionic fields:
\begin{equation}
\lb{defpsi}
\begin{pmatrix}\psi_{1}\\
\psi_{2}
\end{pmatrix}=T\,\begin{pmatrix}\phi_{1}\\
\phi_{2}
\end{pmatrix},
\end{equation}
$H_{int}$ can be rewritten as:
\bea
H_{int} & =&  -\sum_{\bq,lm} \hat g_{lm} \phi_{l,\bq}^{\dagger}\phi_{m,\bq}=\nn\\
&=&-\sum_\bq \left(\L_1 \psi_{1,\bq}^\dagger
\psi_{1,\bq}-\L_2 \psi_{2,\bq}^\dagger \psi_{2,\bq}\right).
\eea
Once defined the new combinations of fermionic fields\cite{footnote_T} $\psi_i$ we
can use the HS decoupling \pref{hs} to write the following partition
function:
\bea
Z&=&\int\mathcal{D}c_{\sigma}^{l}\mathcal{D}c_{\sigma}^{l^{\dagger}}\mathcal{D}h_{i}\mathcal{D}h_{i}^{\dagger}\,
e^{-S},\nn\\
S&=&S_0+\sum_q 
\frac{|h_{1,q}|^{2}}{\Lambda_{1}}+\frac{|h_{2,q}|^{2}}{\Lambda_{2}}\nn\\
\lb{snew}
& &-\sum_q\left(h_{1,q}^{^{*}}\psi_{1,q}+h.c.\right)-i\left(h_{2,q}^{^{*}}\psi_{2,q}+h.c.\right),
\eea
where $q\equiv(i\O_m,\bq)$. 
The action \pref{snew} is now quadratic in the fermionic fields, that can
be integrated out exactly. By introducing the Nambu operators
$N^\dagger_{l,k}=(c^{l\dagger}_{k,\up}, c^{l}_{-k,\down})$ we can indeed
rewrite the action as:
\bea
S&=&\sum_{lk,k'}N_{l,k}^{\dagger}
\left[-\bar G_{k,l}^{-1}\delta_{k,k'}+\Sigma^{l}_{k,k'}\right]N_{l,k'}+\nn\\
&+&\sum_{q}\frac{|h_{1,q}|^{2}}{\Lambda_{1}}+\frac{|h_{2,q}|^{2}}{\Lambda_{2}},
\eea
where:
\begin{eqnarray}
\lb{defgbar}
\bar{G}_{k,l}^{-1} & = & \begin{pmatrix}i\omega_{n}-\xi_{k,l} & T_{1l}\bar
  h_1+iT_{2l}\bar h_{2}\\
T_{1l}\bar
  h_1^*+iT_{2l}\bar h_{2}^* & i\omega_{n}+\xi_{k,l}
\end{pmatrix},\\
\Sigma_{q=k-k'}^{l} & = & \sqrt{\frac{T}{V}}\begin{pmatrix}0 & T_{1l}h_{1,q}+iT_{2l}h_{2,q}\\
T_{1l}h_{1,q}^{^{*}}+iT_{2l}h_{2,q}^{^{*}} & 0
\end{pmatrix}.\nn\\
\lb{defsigma}
\end{eqnarray}
In Eq.\ \pref{defgbar} we put $\bar h_i=\sqrt{T/V}h_{i,0}$. By integrating
out the fermions one gets as usual a contribution to the action equal to
$-\ln det (\bar G^{-1}-\Sigma)=-Tr \ln (\bar G^{-1}-\Sigma)=-Tr \ln  \bar
G^{-1}-Tr\ln (1-\bar G\Sigma)$, where the trace acts both on momentum and
Nambu space. One can then separate the mean-field action
from the fluctuating part as:
\bea
S&=&S_{MF}+S_{FL}\\
\lb{smf}
S_{MF}&=&\frac{\bar h_1^2}{\L_1}+\frac{\bar h_2^2}{\L_2}- \sum_l Tr \ln \bar
G^{-1}_l\\
\lb{sfl}
 S_{FL}&=&
 \sum_{q}\frac{|h_{1,q}|^{2}}{\Lambda_{1}}+\frac{|h_{2,q}|^{2}}{\Lambda_{2}}
 +\sum_l \sum_n  Tr \frac{(\bar G_l\Sigma)^n}{n}
\eea 
From Eq.\ \pref{defgbar} one can see that the HS fields play the role of the SC
gaps in each band, provided that one assumes a saddle-point value of the
antibonding field such that 
\be
\lb{prol}
\bar h_2=iA, \bar h_2^*=i A,
\ee
 to guarantee the Hermitian form of
the saddle-point action. Here we will use instead the generalized transformation
\pref{defpsi} to impose $\bar h_2=\bar h_2^*=0$ at the saddle point. This can be
understood by minimizing the mean-field action \pref{smf}, which gives the
set of equations:
\begin{equation}
\lb{speqs}
\begin{pmatrix}\frac{1}{\Lambda_{1}}-\sum_{l}T_{1l}^{2}\Pi_{l} & -\sum_{l}T_{1l}\Pi_{l}T_{2l}\\
-\sum_{l}T_{2l}\Pi_{l}T_{1l} & -\frac{1}{\Lambda_{2}}-\sum_{l}T_{2l}^{2}\Pi_{l}
\end{pmatrix}\begin{pmatrix}\bar{h}_{1}\\
i\bar{h}_{2}
\end{pmatrix}=0,
\end{equation}
where we defined  the Cooper bubble $\Pi_l$ as:
\be
\lb{defcoop}
\Pi_{l}=\frac{T}{V}\underset{\bk,n}{\sum}\frac{1}{\omega_{n}^{2}+E_{\bk,l}^{2}},
\ee
with the identification
$E_{\bk,l}^{2}=\xi_{\bk,l}^{2}+\left(T_{1l}^{2}\bar{h}^*_{1}\bar
  h_1-T_{2l}^{2}\bar h_2^*\bar h_2+2iT_{1l}T_{2l}(\bar h^*_1\bar h_2+\bar
  h_1\bar h^*_2)\right)$. Once
again this quantity cannot be identified with the energy of the
quasiparticles in each band, unless we use Eq.\ \pref{prol}. 
On the other hand, we can choose the $\varphi$ parameter of the
transformation \pref{grot} to decouple the two saddle-point equations \pref{speqs}:
\be
\lb{fix}
\sum_{l}T_{1l}\Pi_{l}T_{2l}=0.
\ee
In this case one immediately sees that since $\Pi_l>0$ the equation for $\bar h_2$ can only
be satisfied for $\bar h_2=0$, so that the SC transition is only controlled
by the bonding field $\bar h_1$, whose self-consistent equation is
\be
\lb{saddle}
\left[\frac{1}{\Lambda_{1}}-\sum_{l}T_{1l}^{2}\Pi_{l}\right]\bar{h}_{1}=0,
\ee
where we also assumed that $\bar h_1$ is real. This choice corresponds to
the gauge where both gaps are real, as given by (see Eq.\ \pref{defgbar}):
\be
\lb{mfgap}
\Delta_l=T_{1l}\bar h_1.
\ee
We stress once more that even if the saddle-point value of
the antibonding HS field $h_2$  vanishes both gaps are in general
different from zero, and their relative strenght or temperature
dependence is controlled
by the microscopic couplings via the elements of the $T$ matrix. 
The possibility to describe the SC state as a function of a single order
parameter reflects the fact that at $T_c$ only one SC channel becomes
active. To make the connection with a more standard notation, we 
observe that the matrix $T$ in practice diagonalizes the multiband
self-consistency equation, that is usually written as:
\be
\lb{usual}
(\hat g^{-1}-\hat \Pi)\vec{\Delta}=0,
\ee
where $\hat \Pi_{ij}=\d_{ij}\Pi_i$ and $\vec\D$ is a vector formed by the gaps
$\D_l$ in each band. The above equation admits a
non-zero solution $\vec\D$ when the determinant vanishes, i.e. when  (at
least) one eigenvalue is zero. By means of the relations \pref{grot}
above, we see that the set of equations \pref{speqs} and \pref{fix}
corresponds to put the matrix $(\hat \Pi-\hat g^{-1})$ in diagonal form
\be\lb{diagt}
\hat \L^{-1} -T\hat \Pi T^T=T(\hat g^{-1}-\hat \Pi)T^T\equiv Y_i\delta_{ij}
\ee
so that the SC state is reached when the element
$Y_1\equiv1/\L_1-\sum_lT^2_{1l}\Pi_l $ corresponding to the bonding
eigenvalue vanishes, leading to Eq.\ \pref{saddle} above. It should
be noticed that in the two-band case, regardless the intra-band or
inter-band dominated nature of the pairing, the eigenvalues of the matrix
\pref{usual} cannot be both zero, unless the interband coupling
vanishes (see discussion in Appendix C). On the other hand, in the three-band case discussed in
Sec. III below the  matrix \pref{diagt} has three eigenvalues: as we shall
see, when only one of them vanishes one is in the usual SC phase, while the
vanishing of a second eigenvalue signals the emergence of a TRSB
phase. Finally, we notice also that the procedure introduced here to describe a
multiband superconductor in terms of a single ordering field can be
applied also to the case of spatially inhomogeneous  superconducting
condensates, whose Ginzburg-Landau expansion near $T_c$ has attracted
some interest in the recent literature.\cite{shanenko_prb12,shanenko_prb13}

\subsection{Collective modes in the SC state}
Within the present formalism the collective modes in the SC state can be
easily obtained by expanding the action \pref{sfl} up to second order in
the HS fields. In the single-band case, where a single HS field is used to
decouple the interaction, one can follow two alternative but
equivalent roots. Indeed, as relevant variables one can use either (i)
the amplitude and the phase (polar
coordinates) or (ii) the real and imaginary part (cartesian coordinates) of the HS field. In
Appendix A we show how to recover the equivalence between the two
approaches. In our
case, where a single HS field condenses at the transition, the second root
is the only available one. On the other hand, when the interaction has a dominant intraband
character one does not need to use the transformation \pref{defpsi} to get
rid of the antibonding field, and one can introduce HF fields associated directly
to the two gaps in each band. In this case the approach (i) can be again
used, as it has been done for example to study the Leggett mode in MgB$_2$
in Ref.\ [\onlinecite{sharapov_epjb02}], and more recently to investigate Leggett
modes across a TRSB transition for intraband-dominating interactions in Refs.\ [\onlinecite{ota_prb11,
  hu_prl12}]. However, this is not the case physically relevant for
pnictides, as we discussed in the introduction. 

From Eq.\ \pref{sfl} one can see that the coefficients of the Gaussian
action for the HS field will be given in general by BCS correlation functions
computed with the mean-field Green's functions \pref{defgbar}, with the
identification \pref{mfgap} of the band gaps. Following the straightforward procedure
described in Appendix A one then finds that
\bea
\lb{sflg}
S_{FL}&=&\sum_q \eta^T_{-q} \hat S_{FL} (q) \eta_q,\\
\lb{defeta}
\eta^T_q&=&(Re h_{1,q}, iRe h_{2,q}, Im h_{1,q}, iIm h_2{q}).
\eea
Notice that once fixed $\bar h_1$ as real, see Eq.\ \pref{mfgap}, 
one can identify the real
and imaginary parts of the $h_1$  fluctuations as the leading orders in the
amplitude and phase fluctuations of the field, respectively:
\be
\lb{realimag}
Re h_{1,q}=|h_{1,q}|, \quad Im h_{1,q}=\bar h_1 \theta_{1,q}.
\ee
While the same identification cannot be done for the $h_2$ field, we can
still associates its fluctuations to the real and imaginary parts of the
gap fluctuations in each band. Indeed, by following the
same root described in Appendix A to derive the relation between the
averages of the HS fields and the averages of the physical fermionic
operators \pref{defpsi}, one can show that:
\bea
& &\langle \psi_{1,q}\rangle = \frac{1}{\L_1} \langle h_{1,q} \rangle, \quad
\langle \psi_{2,q}\rangle = -\frac{1}{\L_2} \langle ih_{2,q} \rangle, \\
& &\langle \psi^*_{1,q}\psi_{1,-q}\rangle =\frac{1}{\L_1^2}\left( \langle h^*_{1,q} h_{1,-q}\rangle-\L_1\right),\\
& &\langle \psi^*_{2,q}\psi_{2,-q}\rangle =\frac{1}{\L_2^2}\left( \langle
  ih^*_{2,q} ih_{2,-q}\rangle+\L_2\right).
\eea
Since the gap operators in each bands are given by
$\Delta_l=g_{lm}\phi_m=T^T_{lm} \hat\L_{mn}\psi_n$ one can also express the
average values of the gap fluctuations in terms of fluctuations of the
$h_i$ HS fields as:
\bea
\lb{dreal}
& &\langle \Delta_{l,q}+\Delta^*_{l,q} \rangle=T_{1l} \langle Re h_{1,q} \rangle +T_{2l}
\langle iRe h_{2,q} \rangle,\\
\lb{dimag}
& &\langle \Delta_{l,q}-\Delta^*_{l,q} \rangle = T_{1l} \langle Im h_{1,q} \rangle +iT_{2l}
\langle iIm h_{2,q} \rangle,
\eea
and analogous expressions for the correlations functions. As a consequence, we included
the imaginary unit in the $h_2$ components of the fluctuating vector
\pref{defeta} and we will refer in what follows to the first two components
of $\eta_q$ as ``amplitude'' fluctuations and to the last two as ``phase''
fluctuations. Such a decomposition allows one also to easily identify the
character of the fermionic bubbles which appear in the Gaussian
action. Indeed, from Eq.\ \pref{defsigma} one sees that amplitude
fluctuations are associated to a $\sigma_1$ Pauli matrix in the Nambu
notation, while phase fluctuations to $\sigma_2$ (see also Eq.\
\pref{defsigma1} in Appendix A). Moreover, as 
shown in Appendix A, at long-wavelength the amplitude
and phase sectors decouple, so that they are described respectively by
the following $2\times 2$ matrices:
\bea
\hat S^A_{FL}(q)&=& T \hat \L^{11} T^T/2+\hat \L^{-1}=\nn\\
&=&
\begin{pmatrix}
\frac{1}{2}\sum_{l}\Lambda_{l}^{11}(q)T_{1l}^{2}+\frac{1}{\Lambda_{1}} & \frac{1}{2}\sum_{l}\Lambda_{l}^{11}(q)T_{1l}T_{2l}\\
\frac{1}{2}\sum_{l}\Lambda_{l}^{11}(q)T_{1l}T_{2l} &
\frac{1}{2}\sum_{l}\Lambda_{l}^{11}(q)T_{2l}^{2}-\frac{1}{\Lambda_{2}} 
\end{pmatrix}\nn \\
\lb{sampl}
\eea
\bea
\hat S^P_{FL}(q)&=& T \hat \L^{22} T^T/2+\hat \L^{-1}=\nn\\
&=&
\begin{pmatrix}
\frac{1}{2}\sum_{l}\Lambda_{l}^{22}(q)T_{1l}^{2}+\frac{1}{\Lambda_{1}}
& 
 \frac{1}{2}\sum_{l}\Lambda_{l}^{22}(q)T_{1l}T_{2l}\\
 \frac{1}{2}\sum_{l}\Lambda_{l}^{22}(q)T_{1l}T_{2l} &
\frac{1}{2}\sum_{l}\Lambda_{l}^{22}(q)T_{2l}^{2}-\frac{1}{\Lambda_{2}}
\end{pmatrix}\nn\\
\lb{sphase}
\eea
where the $\L_l^{ij}$ bubbles are defined in the Appendix A and the
corresponding diagonal matrices are $\hat \L^{ii}_{lm}\equiv
\L^{ii}_{l}\d_{lm}$. In what follows we shall
investigate the possibility that any collective mode is defined in the
two sectors, by having in mind that a mode corresponds to a solution
of the equation $det \hat S_{FL}(\o=m,\bq=0)=0$ with
$m<2\Delta_{min}$, where $\D_{min}$ is the smallest gap. In practice we
are interested in well-defined resonances below the threshold of the
quasiparticle  excitations: thus it is enough to take into account the
real part of the bubbles $\L^{11}_l,\L^{22}_l$ after analytical
continuation $i\O_m\ra \o+i\d$ to real frequencies, since the imaginary
parts vanish at $\o<2\D_{min}$. 
The different behavior of the collective modes will then follows simply
from the different frequency and momentum dependence of these two
bubbles, whose value at $q=0$ is connected to the Cooper bubble
\pref{defcoop}.  Moreover, as it is shown in the Appendix A (Eq.\
\pref{l22lowq}), at small $q$ and low $T$ one can write:
\bea
\lb{l11}
\L^{11}_l(q)&=&-2\Pi_l+A_l\Delta_l^2+{\cal O}(q^2)\\
\lb{l22}
\L^{22}_l(q)&=&-2\Pi_l+\frac{1}{4\D_l^2}\left(\O_m^2\kappa_l+\bq^2 \frac{\rho_{s,l}}{m_l}\right)
\eea
where $\kappa,\rho_s/m$ represent the compressibility and superfluid
density of each band, respectively, and $A_l=\sum_\bk \tanh (\beta
E_\bk/2)/E_\bk^3$ (see Eq.\ \pref{kulik} below). 

By using Eq.\ \pref{l22} one can write down the $q=0$ limit of
the  phase sector \pref{sphase} as:
\be
\lb{sphaseq0}
\hat S^P_{FL}(q=0)=
\begin{pmatrix}
-\sum_{l}\Pi_l T_{1l}^2+ \frac{1}{\Lambda_{1}}
&  0\\
0 &
-\sum_{l}\Pi_{l} T_{2l}^{2}-\frac{1}{\Lambda_{2}}
\end{pmatrix}
\ee
where we used the constraint \pref{fix} for the $T$ matrix to cancel out
the off-diagonal terms at $q=0$. Eq.\ \pref{sphaseq0} is one of the first
crucial results of the use of the generalized transformation $T$:
indeed, not only it decouples the saddle-point equations, but it also
decouples the phase fluctuations at long wavelengths, connecting their
masses to the eigenvalues of the saddle-point equations themselves, leading to a
straightforward interpretation of the roles of the HS fields. Indeed, since
below $T_c$ $\bar h_1=0$, the self-consistent equation
\pref{saddle} implies that the quantity in square brackets vanishes,
so that one immediately sees that $Im h_{1,q}$ fluctuations describe a massless
mode.  This is not surprising, since from Eq.s\ \pref{mfgap}, \pref{realimag} and \pref{dimag} one
sees that a phase fluctuation for the ordering $h_1$ field corresponds
to a simultaneous change of the overall SC phase in all the gaps:
\be
\lb{defBA}
\Delta_l+ i T_{1l}Im h_1=T_{1l}\bar h_1 +iT_{1l}\bar h_1 \theta_1\simeq \Delta_l
e^{i\theta_1}.
\ee
As a consequence, $Im h_1$ is the Goldstone mode of the SC transition,
that is expected to be massless in the SC phase. For what concerns
instead the fluctuations of the antibonding field $h_2$ we can first
analyze the small frequency expansion of Eq.\ \pref{sphase} that follows
from Eq.\ \pref{l22}, i.e.:
\be
\lb{sphasesmall}
\small{
\hat S^P_{FL}=-
\begin{pmatrix}
\frac{\o^2}{8}\sum_l \kappa_l \frac{T_{1l}^{2}}{4\D_l^2} & \frac{ \o^2}{8}\sum_l
\kappa_l \frac{T_{1l} T_{2l}}{4\D_l^2} \\  
\frac{ \o^2}{8}\sum_l \kappa_l \frac{T_{1l} T_{2l}}{4\D_l^2} &
\frac{ \o^2}{8}\sum_l \kappa_l \frac{T_{2l}^{2}}{4\D_l^2}  +\left(\sum_l
  \Pi_l T_{2l}^{2}+\frac{1}{\L_2}\right)
\end{pmatrix}},
\ee
where the analytic continuation $i\O_m\ra 
\o+i\d$ has been made. As $\o\ra 0$ one sees that $Im
h_1$ and $Im h_2$ decouple, and one recovers the massless $Im h_1$ mode, as
discussed above.  On the other hand,
the fluctuations of the antibonding $h_2$ field do not give rise to any
collective mode. Indeed, the $22$ element of the matrix
\pref{sphasesmall} does not admit any real solution for $\o$, due to
the fact that the quantity in brackets is strictly positive. This
result, which is confirmed by the explicit calculation of
$\L_l^{22}(\o)$ at all frequencies and temperatures, is a direct consequence of the fact that the $h_2$ field is
associated to the antibonding SC channel of the system.  Indeed, as we show in
details in the Appendix B, if $h_2$ were associated to a bonding SC channel
(i.e. a positive eigenvalue in Eq.\ \pref{gdiag}), the $-1/\Lambda_2$
term in Eq.\ \pref{sphase} would be replaced by $+1/\Lambda_2$,
leading to a well-defined mode in Eq.\ \pref{sphasesmall}, that coincides with the 
usual Leggett mode, see Eq.\ \pref{oml}.  It is also worth stressing that the {\em absence}
of the Leggett mode in a two-band modelization of pnictides does not
mean that relative phase fluctuations of the gaps in the two bands are
absent, but simply that these fluctuations do not define a coherent
collective mode of the system.

\begin{figure}[htb]
\includegraphics[scale=0.3, angle=0,clip=]{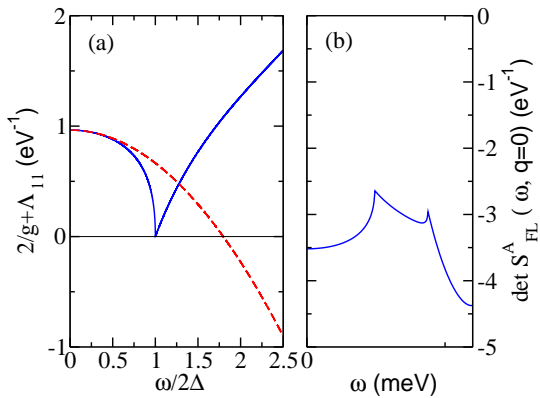}
\caption{(color online) Left panel: frequency dependence of the
  amplitude fluctuations as given by Eq.\ \pref{kulik} in the
  single-band case. The vanishing at $\o=2\Delta$ signals the presence
  of an amplitude mode with mass $m=2\Delta$. Notice that the 
quadratic low-frequency expansion, given by the dashed line, would
lead to a wrong mass $m\simeq \sqrt{12}\D$. Right panel:
determinant of the action \pref{sampl} at $T=0$ in the amplitude sector as a
function of $\o$ for coupling values $N_1=1$ eV$^{-1}$, $N_2=2.2$
eV$^{-1}$, $g_{12}=0.5$ eV, $\o_0=15$ meV.  The
determinant never vanishes, so that no well-defined mode is found in
this case. The overall negative sign is due to the presence of the
antibonding channel, see also Eq.\ \pref{samplsmall}.}
\label{fig_l11}
\end{figure}

For what concerns instead the amplitude sector \pref{sampl}, by using
again the self-consistent equations
\pref{fix}-\pref{saddle} and the relation \pref{l11} one sees that at
$\bq =0$ in general 
\be
\lb{samplsmall}
\hat S^A_{FL}=
\begin{pmatrix}
A +{\cal O}(\o^2)  & B+{\cal O}(\o^2)\\
B+{\cal O}(\o^2) &
-C+{\cal O}(\o^2)
\end{pmatrix} \\
\ee
where $A=\frac{1}{2}\sum_{l}A_l \D_l^2T_{1l}^{2}$,
$B=\frac{1}{2}\sum_{l} A_l \D_l^2 T_{1l}T_{2l}$, $C=\sum_{l} C_l
  T_{2l}^{2}+\frac{1}{\Lambda_{2}}$ are positive constants, with
  $C_l=\sum_\bk (\xi_\bk^2/E_\bk^3)\tanh (\beta E_\bk/2)$. As one could expect, there is no
  massless mode in the amplitude sector, since amplitude fluctuations
  are always costly in the SC phase. One could then wonder if massive
  modes are present. In the
single-band case one knows that amplitude fluctuations at $\bq=0$
correspond to a well-defined mode with frequency $m=2\Delta$, which
get easily damped by interactions\cite{volkov_zet73, schimd_prl75}. This result
follow from the fact that the coefficient of the amplitude
fluctuations reduces (see Eq.\ \pref{sflc}) to ($g$
being the SC coupling)\cite{kulik_jltp81}
\bea
& &\Lambda^{11}(\o,\bq=0)+\frac{2}{g}=\nn\\
&=&\sum_\bk
\tanh\left(\frac{E_{\mathbf{k}}}{2T}\right)\left[-\frac{\xi_{\mathbf{k}}^{2}}{E_{\mathbf{k}}^{2}}\left(\frac{1}{2E_{\mathbf{k}}
+\omega}+\frac{1}{2E_{\mathbf{k}}-\omega}\right)+\frac{1}{E_{\mathbf{k}}}\right].\nn\\
\lb{kulik}
\eea
This function of $\o$ vanishes at $\o=2\Delta$ with a square-root
singularity,\cite{kulik_jltp81} and it is positive everywhere else, see Fig.\
\ref{fig_l11}a. In the multiband case described by Eq.\ \pref{sphase}
above one is then mixing the $\L_l^{11}$ bubbles of the two bands,
which have in general zeros for two different values $2\D_l$. For
this reason, unless one considers strictly identical bands, the
$det \hat S^P_{FL}(\o,\bq=0)$ never vanishes, as shown in Fig.\
\ref{fig_l11}b, so that well-defined amplitude modes are
absent. This example shows also that one should be very careful in
computing the collective modes by making a low-energy expansion of the
$\L^{11}_l\simeq -2\Pi_l+A_l\Delta^2-B_l2\o^2$
bubbles.\cite{stanev_prb12,maiti_prb13}
Indeed, one could obtain either
spurious results or masses which are quantitatively wrong, especially
in the TRSB phase where amplitude and phase fluctuations get
mixed.\cite{babaev_prb11,stanev_prb12,maiti_prb13}  We will
come back to this point at the end of the next Section.

Finally, we observe that above $T_c$ the phase and amplitude sectors
become degenerate, as expected, and one recovers the results discussed
in Ref.\ [\onlinecite{fanfarillo_prb09}]. Indeed, the $\L_l^{11}$ and
$\L_l^{22}$ bubbles coincide, and the leading terms at small $q$ go
like $\eta \bq^2, \gamma |\O_m|$. 
More specifically, we observe that at $q=0$
the action for the Gaussian fluctuations coincides with the usual quadratic
expansion of the free energy, and it s given by:
\bea
S_{FL}(q=0)&=&\left
  [\frac{1}{\L_1}-\sum_lT_{1l}^2\Pi_l\right]|h_1|^2+\nn\\
&+&\left [\frac{1}{\L_2}+\sum_lT_{2l}^2\Pi_l\right]|h_2|^2.
\eea
As one can see the coefficient of the $h_2$ field is always {\em positive},
showing that it never orders. In contrast, a wrong application of the HS
transformation \pref{hs} lead the authors of
Refs. [\onlinecite{maiti_prb13,fernandes_cm13}] to the counterintuitive
result that the coefficient of the antibonding field is always {\em
  negative}, making it difficult to justify why it should not order. This
shows once more that an extra care is needed to extend to
interband-dominated interactions the results known for single-band systems,
where a single bonding SC channel exists.

\section{Three-band model for the TRSB transition}
\label{3bands}

\subsection{Occurrence of a TRSB state in pnictides}

\begin{figure}[htb]
\includegraphics[scale=0.35, angle=0,clip=true]{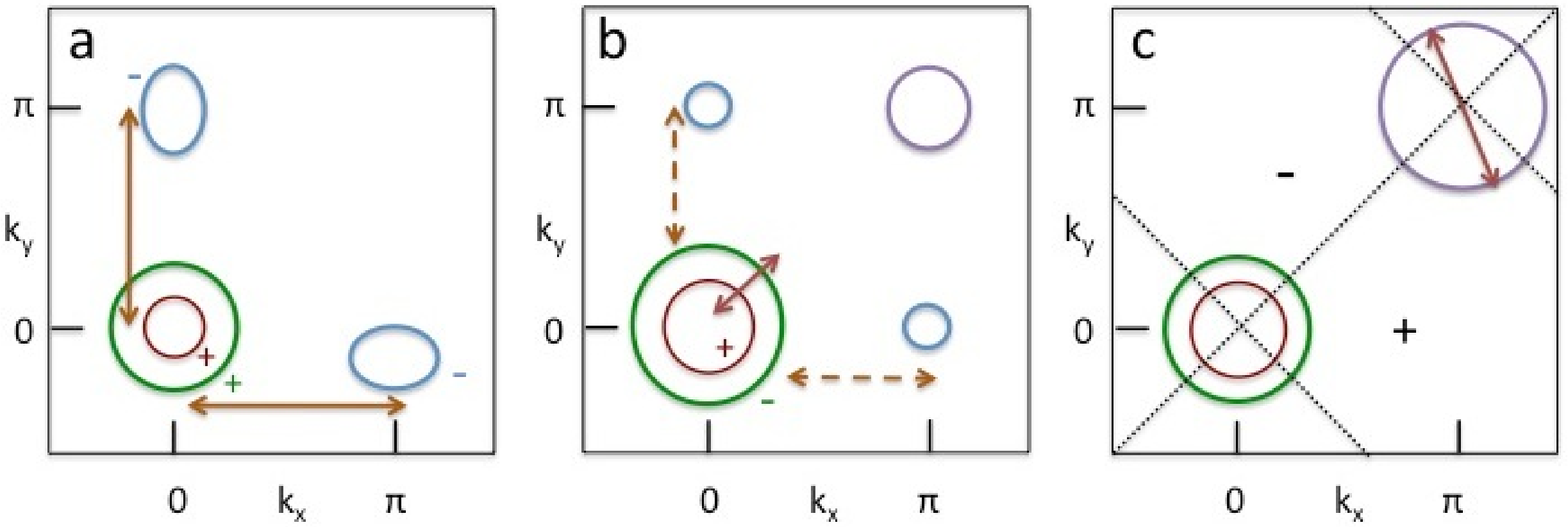}
\caption{(color online) Schematic of the band structures in pnictides
  in the unfolded Brollouin zone.  (a) Typical band structure for optimally-doped 122
  compounds (like e.g. Ba$_{0.4}$K$_{0.6}$Fe$_2$As$_2$), formed by two
  hole pockets around $\G$ and two electron ones at $(0,\pi)$ and
  $(\pi,0)$. In this case the largest coupling is an interband
  repulsion between the hole and electron Fermi sheets, leading to the
  $s_\pm$ symmetry of the order parameter with a sign change of the
  gap between hole and electron bands. (b)  Strong hole doping: in
  this case the electron pockets reduce considerably and a third hole
  pocket appears at $(\pi,\pi)$.  In the case of KFe$_2$As$_2$ (c) the
  electron pockets disappear completely. It has been argued that $s$-wave (b) and
  $d$-wave (c) symmetries are nearby in energy at strong hole doping. In the $s$-wave symmetry
the change of sign of the gaps occurs between the two hole pockets at
$\G$, while on the remaining bands the order parameter is very small. In the
$d$-wave symmetry (c) instead the largest gap is on the third hole pocket
and nodes are present on all the Fermi surfaces.}
\label{fig-scheme}
\end{figure}

Once established the general properties of the collective modes in a
superconductor where bonding and antibonding SC channels coexist, let us now focus
more specifically on the case of a three-band model for pnictides,
where an additional repulsion between the two hole bands is
considered. This case has attracted considerable interest in the
recent literature due to the experimental advances in making 122 samples\cite{takahashi_prb11,shin_prb12,matsuda_cm13}
heavily hole-doped away from half-filling, until the end member
KFe$_2$As$_2$ is
reached.\cite{hashimoto_prb10,taillefer_prl12,okazaki_science12,taillefer_natphys13,matsuda_cm13}
Even though a full agreement between
theoretical predictions and experimental results has not been reached yet, we
would like to summarize here some results relevant for the focus of
the present manuscript. A schematic of the band-structure evolution
from Ba$_{1-x}$K$_x$Fe$_2$As$_2$ to KFe$_2$As$_2$ in the
unfolded Brillouin zone (one Fe atom per unit cell) is shown in Fig.\
\ref{fig-scheme}. At intermediate doping (Fig.\ \ref{fig-scheme}a) the
system admits two hole pockets at $\G=(0,0)$ and two electron pockets
at $(\pi,0)$ and $(0,\pi)$. The largest interactions in this situation
are the spin-fluctuations mediated inter-pocket repulsions between hole
and electron bands,\cite{hirschfeld_njop09,chubukov_prb11} that lead to the $s_\pm$ symmetry of the order
parameter, i.e. constant gaps on all the FS with a change of sign
between hole and electron bands. In this situation, by neglecting
nematic effects making the electron pockets inequivalent,  an
effective two-band description as the one discussed in the previous
section is possible.\cite{benfatto_prb08,fanfarillo_prb09} As doping increases the electron pockets shrink
and a third hole pocket around $(\pi,\pi)$ appears (Fig.\
\ref{fig-scheme}b), until that only hole pockets remain for
KFe$_2$As$_2$ (Fig.\ \ref{fig-scheme}c). In this compound several
theoretical
calculations\cite{thomale_prl11,suzuki_prb11,chubukov_prb11} have
shown that $s$-wave and $d$-wave symmetry are almost degenerate in
energy. However, the gap hierarchy would be very different in the two
cases: in the $s$-wave case the leading interaction is an 
interband repulsion at small momentum between the hole pockets at
$\G$, so that the sign change between the gaps is now realized between
the two central hole bands (having eventually accidental
nodes\cite{chubukov_prb12}), while on the remaining pockets the gap is
vanishing. Instead the  $d$-wave symmetry  is driven by a large
intraband repulsion within the hole pocket at $(\pi,\pi)$, so that
the gap is largest here and nodes are present on all the FS. The
experimental situation is quite controversial: while ARPES
measurements show no nodes at large ($x=0.7$) K
doping\cite{takahashi_prb11,shin_prb12} or accidental nodes for
KFe$_2$As$_2$,\cite{okazaki_science12} thermal probes of the
quasiparticle excitations indicate nodal
gaps.\cite{hashimoto_prb10,taillefer_prl12} 

From the point of view of
the general description of the collective modes that we will give here,
the relevant aspect is that once that two SC channels are almost
degenerate in energy one can eventually access a phase where both of
them coexist, leading to a TRSB state. To slightly simplify the
notation and to make contact with previous work on this topic\cite{stanev_prb10,stanev_prb12,maiti_prb13} we will
discuss here the case where the order parameter remains in the
$s$-wave symmetry class, so that the most relevant interactions are
interband repulsion between hole and electron bands, and within the
hole pockets at $\G$. By assuming again degenerate electron pockets one
can then investigate for example the minimal three-band model  proposed in
Ref.[\onlinecite{stanev_prb10}] where the two hole bands (bands 1,2) are equal, so
that the matrix
$\hat g$ of Eq.\ \pref{defg2} becomes for this three-band case:
\be
\lb{defg3}
\hat g=-\begin{pmatrix}0 & V_{hh} & V_{he}\\
V_{hh} & 0 & V_{he}\\
V_{he} & V_{he} & 0
\end{pmatrix}
\ee

\begin{figure}[htb]
\includegraphics[scale=0.3, angle=0,clip=]{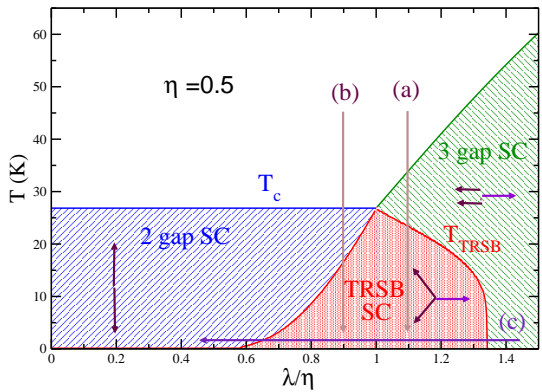}
\caption{(color online) Phase diagram of the model \pref{defg3}
  obtained by numerical solution of the mean-field equations
  \pref{usual}. Here we used as bosonic scale
  $\o_0=15$ meV. Notice that the $T_{TRSB}$ line ends
  at a finite value of $\lambda/\eta$ for $\l>\eta$, while for
  $\l<\eta$ $T_{TRSB}$ is in principle always finite but it is
  exponentially suppressed as one moves to small $\l$ values.}
\label{fig-pd}
\end{figure}

As it has been noticed in Ref.\ [\onlinecite{stanev_prb10}], despite the
fact that the mean-field equations in this three-band model appear
as a straightforward generalization of the two-band case discussed in
the previous Section, the intrinsic frustration hidden in the SC model
\pref{defg3} leads to the appearance of a qualitatively new effect,
i.e. the possible emergence of a $s+is$ state which breaks time-reversal
symmetry. Indeed, in the model \pref{defg3} each band would like to
have a gap of opposite sign with respect to the gap in the other
bands: when three gaps compete one then realizes a situation analogous
to the antiferromagnet in the triangular lattice, where spins orients
themselves at relative $2\pi/3$ angles. In the SC problem the
frustration occurs in the range of parameters (i.e. interactions and/or
temperature) where two eigenvalues of the matrix $\hat \Pi-\hat
g^{-1}$ vanish (see Sec. III-B and Appendix C), allowing for an intrinsically complex  SC order
parameter. Since in this case
$\Delta_l^*\neq \Delta_l$ time reversal (which corresponds to complex
conjugation) is spontanously broken and the system is in a $s+is$ TRSB
state. To give a general idea of the range of parameters for the TRSB
phase we  show in Fig.\ \ref{fig-pd} the phase diagram of the specific
model \pref{defg3}. Here we assumed for simplicity that the DOS
$N_l\equiv N$ in
all bands are equal, so that we can introduce the dimensionless
couplings 
\be
\lb{defetal}
\eta=V_{hh}N, \quad \l=V_{he}N.
\ee
As it has been discussed previously\cite{stanev_prb10,maiti_prb13},
the $T_{TRSB}$ separating the normal superconductor from the TRSB
state ends at a finite value $\l_{cr}(\eta)>\eta$, while for $\l<\eta$ the
TRSB phase is always present at $T=0$, but the $T_{TRSB}$ is
exponentially suppressed. By using $\eta=0.5$ and 
$\o_0=15$ meV for the BCS bosonic scale in the  $\Pi_l$ bubbles,
as roughly appropriate for pnictides,\cite{benfatto_prb09} 
one obtains $\l_{cr}\simeq 1.3$, leading to  a reasonable wide range of parameters where the TRSB
transition can occur.  Even though these numbers have to be considered only
indicative for real materials, due to the simplifications of the model
\pref{defg3} and the overestimation of the critical temperatures in
mean-field like calculations, for a specific sample one could indeed
observe one of the thermal transitions marked by vertical lines in
Fig.\ \ref{fig-pd}. 

As we discussed above, several other possibilities exist for an 
intermediate TRSB state in pnictides, depending on the nature of the
competing SC channels, that reflects on the structure of the
matrix \pref{defg3} and on its eigenvalues. For example, to account
for a possible $d$-wave symmetry in 
KFe$_2$As$_2$ one should include a third hole pockets with
a large intraband-repulsion
term.\cite{chubukov_prb11} On the
other hand, the same matrix structure
\pref{defg3} but with a
different identification of the bands\cite{fernandes_cm13} can be used to describe the
$s+id$ state that has been proposed for the electron-doped pnictides.
\cite{wu_prl09,thomale_prb12,fernandes_prl13,fernandes_cm13} In this case,
by adding to the schematic structure of Fig.\ \ref{fig-scheme}a an
interband repulsion between the electron pockets\cite{thomale_prb12,fernandes_cm13} one could induce a
SC state with a sign change of the gap between the $(0,\pi)$ and
$(\pi,0)$ pockets, that corresponds to $d$-wave symmetry, even if
without nodes on the FSs. As we shall see below, our approach allows
us to establish a general correspondence between the structure of
mean-field equations \pref{usual} and the evolution of the collective modes, providing
thus a general scheme to test experimentally whether or not a TRSB state is realized, regardless the specific
symmetry of the two degenerate SC channels active in the TRSB phase. For this reason,
while previous work has focused on the $T=0$ behavior of the
collective modes as a function of the tuning parameter (i.e. the
doping) for the {\em quantum} TRSB transition, here we focus on the
possibility to identify the occurrence of a {\em thermal} TRSB
transition. Indeed, while the quantum phase transition between the TRS
and TRSB state has in general only two end points, the thermal
transition occurs in a much wider range of parameters, making
eventually its identification a  more accessible experimental task. 

\subsection{Collective mode across the TRSB transition}
To extend the collective-modes derivation of Sec.\ \ref{2bands}  we
will start by considering the most general three-band model which
admits a TRSB state and has one antibonding SC channel. Thus, $\hat g$ must have
two positive and one negative eigenvalue, so that after the rotation $R$ the
diagonal matrix $\hat\L$ is:
\be
\lb{g3diag}
\hat g=R^{-1}\begin{pmatrix}\Lambda_{1} & 0 & 0\\
0 & \Lambda_{2} & 0\\
0 & 0 & -\Lambda_{3}
\end{pmatrix}R,
\ee
where we also set by definition $\L_1\geq \L_2$. For example, in the simplified model
\pref{defg3} one has that for $\eta>\l$ is $\L_1=\eta$,
$\L_2=(\sqrt{\eta^2+8\l^2}-\eta)/2$ and
$\L_3=(\sqrt{\eta^2+8\l^2}+\eta)/2$, while for $\eta<\l$ the role of
$\L_1$ and $\L_2$ is interchanged. The derivation of the effective
action is then a straightforward generalization of the procedure used
in Sec.\ \ref{2bands}. In particular, also in this case one has to introduce a HS field $h_3$
associated with the repulsive channel $\L_3$, and one can take advantage
of the generalized transformation $T$ (depending now on three parameters,
see Appendix C) to impose $\bar h_3=0$ at the
saddle point. 
Indeed, the equivalent of the saddle-point Eq.s\ \pref{speqs}
can be made diagonal again by using the three conditions which generalize
Eq.\ \pref{fix}, i.e.
\be
\lb{fix3}
\sum_{l}T_{il}\Pi_{l}T_{j\neq i,l}=0, \quad i,j=1,2,3,
\ee
so that one is left with 
\bea
&&{\small 
\begin{pmatrix}\frac{1}{\Lambda_{1}}-\sum_lT_{1l}^{2}\Pi_{l} & 0 & 0\\
0 & \frac{1}{\Lambda_{2}}-\sum_lT_{2l}^{2}\Pi_{l} & 0\\
0 & 0 & -\frac{1}{\Lambda_{3}}-\sum_lT_{3l}^{2}\Pi_{l}
\end{pmatrix}\begin{pmatrix}\bar{h}_{1}\\
\bar{h}_{2}\\
i\bar{h}_{3}
\end{pmatrix}=}\nn\\
\lb{speqs3}
&=&Y_1 \bar h_1 + Y_2 \bar h_2 + iY_3 \bar h_3=0,
\eea
where the $Y_i$ are the eigenvalues of the matrix $\hat g^{-1}-\hat \Pi$
which enters the usual mean-field equations \pref{usual}. 
As one can see, in full analogy with the two-band case \pref{speqs}
above, the coefficient $Y_3$ which multiplies the antibonding HS field
$\bar h_3$ is always positive, so that one imposes $\bar h_3=0$ at the
saddle point. The remaining two coefficients $Y_1,Y_2$ can both in
principle vanishing, leading to finite saddle-point values of the
corresponding HS fields. 

Let us discuss the thermal evolution equivalent to one of the paths (a),(b)
in Fig.\ \ref{fig-pd}, starting from the non-SC state.  As $T$ decreases
and the Cooper bubbles increase the first coefficient which vanish at $T_c$
in Eq.\ \pref{speqs3} is for example $Y_1$.  Then $h_1$ is the
first HS field which orders. Its phase can be chosen real 
$\bar h_1=R_1$ real, so all the gaps are given by Eq.\
\pref{mfgap} and are real. As the temperature decreases further, 
according to the range of parameters of the matrix $g$, it is possible
that at $T=T_{TRSB}$ also $Y_2(T_{TRSB})$ vanishes, 
\be
\lb{ttrsb}
Y_2(T_{TRSB})=\sum_l T^2_{2l}(T_{TRSB})\Pi_l(T_{TRSB})-1/\L_2=0.
\ee
In this case, as we discuss in the Appendix C, one can also show that at
lower temperatures the imaginary part of $\bar h_2$ acquires a finite
saddle-point value. More specifically, one can always choose a gauge where 
$\bar h_2$ is purely imaginary, i.e. $\bar h_2=i
I_2$. As a consequence, the mean-field gaps at $T<T_{TRSB}$ are given by
\be
\lb{mfgaptrsb}
\Delta_l=T_{1l}\bar h_1+T_{2l} \bar h_2= T_{1l}R_1+iT_{2l} I_2=|\D_l|e^{i\bar\vartheta_l}
\ee
so that they are intrinsically complex and a TRSB state is
reached. Moreover, the additional $Z_2$ symmetry between the two possible
time-reversal-symmetry breaking ground states \pref{mfgaptrsb}
is encoded in the complex conjugation for the $\bar h_2$ field, that leads
to a change of sign of
all the phases $\vartheta_l$ without changing the ground-state energy. 

The emergence of a finite imaginary part of $\bar h_2$ below TRSB has a
precursor effect on the behavior of the collective phase modes above
$T_{TRSB}$. Indeed, in the TRS phase where all the gaps have trivial phases
one can obtain a straightforward extension of Eq.s\
\pref{sampl}-\pref{sphase}  for the amplitude and phase
fluctuations of the HS fields. In particular, by using again the
constraints \pref{fix3} for the $T$ transformation, the equivalent of
Eq.\ \pref{sphaseq0} for the phase sector $\eta_q^T=(Im h_{1,q}, Im
h_{2,q}, i Im h_{3,q})$ in the
long-wavelength $q\simeq 0$ limit can be written as:
\bea
&&
{\small 
\hat S^P_{FL}(q=0)=}\nn\\
&&
{\small =\begin{pmatrix}
-\sum_{l}\Pi_l T_{1l}^2+ \frac{1}{\Lambda_{1}}
&  0 & 0\\
0 & 
- \sum_{l}\Pi_{l} T_{2l}^{2}+\frac{1}{\Lambda_{2}}  & 0\\
0 & 0 & -\sum_{l}\Pi_{l} T_{3l}^{2}-\frac{1}{\Lambda_{3}}
\end{pmatrix}}\nn\\
\lb{sphase3q0}
\eea
Eq.\ \pref{sphase3q0} is one of the central results of our paper. Indeed, it
establishes a direct correspondence between the masses of the phase
modes and the saddle-point equations \pref{speqs3}, 
showing  that as soon as one reaches the TRSB state, defined by Eq.\ \pref{ttrsb}, the fluctuations of the $Im
h_2$ HS field become {\em massless}.  It must be emphasized that the this
result holds regardless the structure of the coupling matrix. Indeed, one
can prove (Appendix C) that necessary and sufficient condition to have gaps
with non-trivial phases is that two eigenvalues of the matrix $\hat g^{-1}-\hat
\Pi$, i.e. two $Y_i$ coefficients in the diagonal form
\pref{speqs3}, must vanish. Since the $T$ transformation decouples also the
phase modes and connects their masses at $T\geq T_{TRSB}$ to the $Y_i$
coefficients, it makes possible to show in full generality that at the
boundary between a TRS and TRSB phase one additional phase mode becomes
massless. By considering then the phase diagram of Fig.\ \pref{fig-pd}, such a massless
mode emerges along all the line $T_{TRSB}$, as well as for isothermal
transitions\cite{babaev_prb11,hu_prl12,maiti_prb13} as a function of the  coupling parameters for the matrix
$\hat g$, like path (c). In this case the TRSB
state would be equally determined by the condition $Y_2=0$, considering
$Y_2$ a function e.g. of the SC coupling $\l$:
\be
\lb{qttrsb}
Y_2(\l_{TRSB})=\sum_l T^2_{2l}(\l_{TRSB})\Pi_l(\l_{TRSB})-1/\L_2=0
\ee
It is worth stressing that our derivation shows also that in the three-band
case only one
additional mode (other than the Bogoliubov-Anderson Goldstone mode) can be
massless at the TRSB transition. Indeed, for interband-dominated coupling the
fluctuations of the antibonding $h_3$ field in Eq.\ \pref{sphase3q0} do not
identify a mode, as explained in Sec.\ \ref{2bands}. On the other hand, if 
also the third eigenvalue $\L_3$ of the matrix \pref{g3diag} were positive,
the associated $h_3$ fluctuations would describe a Leggett-like mode that
cannot become massless, since at least one eigenvalue of the decomposition
\pref{speqs3} must be finite (see Appendix C). The possibility to establish these
results on general grounds is crucial to identify the total number of
massless modes a-priori. Indeed, an explicit numerical calculations of the
collective modes, done e.g. by using the low-frequency expansion of
the bubbles,\cite{stanev_prb12,maiti_prb13} becomes very delicate when one of the gap vanishes, as we shall
discuss in more details in the next Section. 

Below $T_{TRSB}$ the behavior of the collective modes is more
complex, due to the mixing between amplitude and phase
fluctuations.\cite{babaev_prb11,stanev_prb12,maiti_prb13}  Indeed, when the SC gaps $\Delta_l$ in each band are
complex numbers the fermionic bubbles which appears in Eq.s\
\pref{sampl}-\pref{sphase} acquire an explicit dependence
on the saddle-point values $\bar\vartheta_l$ of the phases of the 
SC order parameters. More specifically
one has that 
\bea
\lb{l11t}
\L^{11}_l(q)&=&\bar \L^{22}_l(q)+2\cos^2 \bar \vartheta_l F_l(q),\\
\lb{l22t}
\L^{22}_l(q)&=&\bar \L^{22}_l(q)+2\sin^2 \bar \vartheta_l F_l(q),\\
\lb{l12t}
\L^{12}_l(q)&=&2\sin \bar \vartheta_l\cos\bar\vartheta_l F(q)+{\cal O}(q^2),
\eea
where $\bar \L_l^{22}$ is a function of $|\Delta_l|$, so it coincides with the expression \pref{gen} of the $\L^{22}$
bubble computed assuming a real gap, and 
\be
F_l(q)=2|\D_l|^2\frac{T}{V}\sum_{\bk,n} \frac{1}{(\O_m+\o_n)^2+E^2_{\bk+\bq}}\frac{1}{\o_n^2+E^2_{\bk}}
\ee
is also a function only of the gap amplitude $|\Delta_l|$.  When the gaps
have trivial phases $\vartheta_l=0,\pi$ these definitions coincide with the
ones given in Appendix A and one recovers the expansion
\pref{l11}-\pref{l22} used above. Below $T_{TRSB}$ the most important
difference is that the bubbles $\L^{12}_l$ which
appear in the coupling between the amplitude and phase sectors (see Eq.\
\pref{sflc}) cannot be neglected,  making the structure of the Gaussian
fluctuations \pref{sflg} considerably more complicated. 
In this situation the structure of the collective modes is
not simplified by the use of the transformation $T$. Thus, in order to
simplify the numerical computation, we will take advantage of the fact that
thanks to Eq.s \pref{sampl}-\pref{sphase} the overall action for mixed
amplitude and phase fluctuations is a six times six matrix given by:
\bea
\hat S_{FL}&=&\frac{1}{2}\begin{pmatrix}
T\hat \L^{11}T^T+2\hat \L^{-1} & T\hat \L^{21}T^T\\
T\hat \L^{12}T^T & T\hat \L^{22}T^T+2\hat \L^{-1}
\end{pmatrix}=\nn\\
\lb{smatrix}
&=&\frac{1}{2}
\hat T \begin{pmatrix}
\L^{11}+2\hat g^{-1} & \hat \L^{21}\\
\hat \L^{12} & \hat \L^{22}+2\hat g^{-1}
\end{pmatrix}\hat T^T\equiv\hat T \hat M \hat T^T
\eea
where we used the property \pref{grot} that $\hat \L^{-1}=Tg^{-1}T^T$
and we defined $\hat T$ as a 6$\times$6 diagonal matrix  having the 3$\times$3
matrix $T$ on the diagonal. Since $det \hat T=1$ the
collective modes will be given by the solutions of the equation $det
\hat M=0$. It is worth noting that the corresponding eigenvectors can
be associated to amplitude and phase fluctuations in the various bands:
indeed, the relations \pref{dreal}-\pref{dimag} between the fermionic
operators and the HS fields will read in this case:
\bea
\lb{newre}
\langle Re \D_{l,q}\rangle &=&T_{l1} \langle Re h_{1,q}\rangle +T_{l2}
\langle Re h_{2,q}\rangle +T_{l3} \langle iRe h_{3,q})\rangle \\
\lb{newim}
\langle Im \D_{l,q}\rangle &=&T_{l1} \langle Im h_{1,q}\rangle +T_{l2}
\langle Im h_{2,q}\rangle +T_{l3} \langle iIm h_{3,q}\rangle
\eea
which correspond in a short notation to e.g. $\langle Re \D\rangle =T^T
\langle Re h\rangle$, with the usual inclusion of the imaginary unit
in the fluctuations of the antibonding field $h_3$. Thus it is not
surprising that the $\hat M$ matrix coincides with the derivation done in Refs.\
[\onlinecite{stanev_prb12,maiti_prb13}] by means of linear response
theory in the band basis. In addition, in the case of dominant intraband
pairing, where no imaginary unit is associated to the HS  fields, the
relations \pref{newre}-\pref{newim} can be used to define new bosonic
variables. In this case,  when all the gaps are opened so that $Im
\Delta_{l,q}=\Delta_l \vartheta_{l,q}$, by means of the identity \pref{l22lowq} 
one recovers for the phase sector the same
structure reported in Ref.\ [\onlinecite{hu_prl12}]. 
Notice also that the coupling between fluctuations in different bands is
provided by the inverse matrix $\hat g^{-1}$ of the SC
couplings, while the coupling between the amplitude and phase sector is
diagonal in the band index and it is given by the $\L^{12}_l$ bubbles of Eq.\
\pref{l12t}, which are proprtional to the $\sin \bar\vartheta_l$, so
that they differ from zero only in the TRSB state. This result s very general, and indeed it can be found also within the
phenomenological multiband Ginzburg-Landau approach of Ref.\
[\onlinecite{babaev_prb11}], where the interband couplings are
provided by Josephson-like terms.  

\subsection{Temperature and coupling dependence of the Leggett mode}

\begin{figure}[htb]
\includegraphics[scale=0.3, angle=0,clip=]{fig4a.eps}
\includegraphics[scale=0.3, angle=0, clip=true]{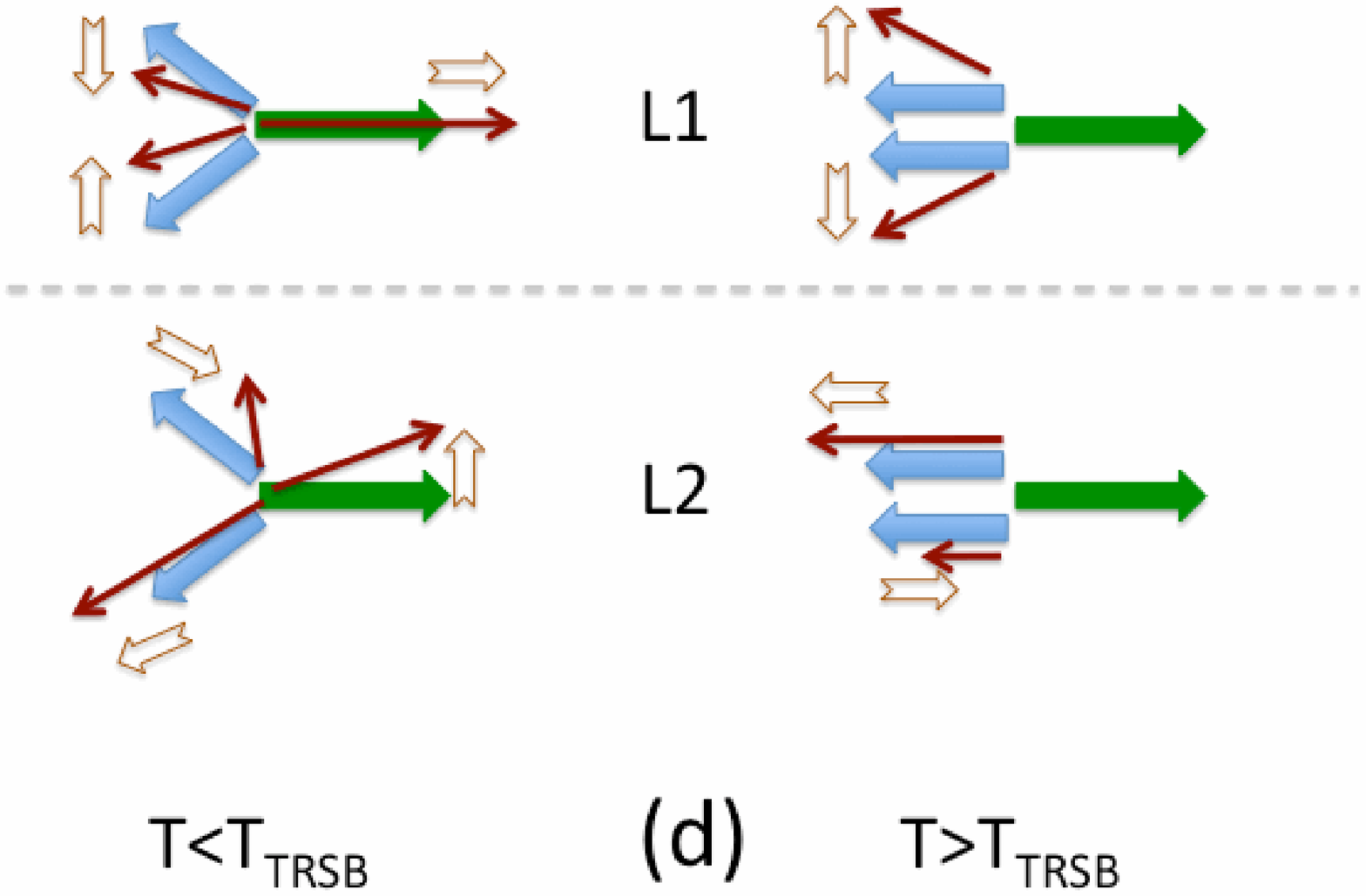}
\caption{(color online) Temperature evolution of the low-energy modes
  along the path (a) of Fig.\ \ref{fig-pd}, corresponding to
  $\l=1.12\eta$. Here $T_c=35.5$ K and $T_{TRSB}=23$ K. (a): Temperature
  dependence of the low-energy mode along with the minimum gap threshold,
  obtained by the temperature dependence of the gaps reported in the inset
  along with the phases of the hole gaps. (b) and (c): components of the
  eigenvectors corresponding to the modes labeled as L1 and L2 in panel
  (a). (d): schematic structure of the modes below and above
  $T_{TRSB}$. Here big full arrows denote the equilibrium gaps while the
  thin arrows denote the gaps including the fluctuations, identified by the big empty arrows. As
  one can see, L1 evolves in the ordinary Legget mode for the order
  parameters in the two hole bands, while L1 evolves towards an amplitude
  mode.}
\label{fig-l112}
\end{figure}

To show explicitly the temperature evolution of the Leggett modes we will
refer for simplicity to the set of coupling constants defined by eq.\
\pref{defg3}, which gives rise to the phase diagram shown in Fig.\
\ref{fig-pd}. 
As one can see, 
while Eq.\ \pref{smatrix} does not allow for a simple identification of the number and nature of
the collective modes, it simplifies the numerical evaluation of the
modes since one does not need to determine also the $T$ matrix. 
We then solved self-consistently the gap
equations and computed the matrix $\hat M$ in Eq.\ \pref{smatrix}, looking
for well-defined modes below the threshold $2\Delta_{min}$ provided by
the smallest gap in the problem. We assume conventionally that
the gap in the electron band $\Delta_3$ is real and positive, while
the gaps in the hole bands are $\Delta_1=\D e^{i\phi_1}$, $\Delta_2=\D
  e^{i\phi_2}$. According to the phase diagram of Fig.\ \ref{fig-pd},
  the three phases correspond respectively to:

\bea
\mathrm {3\, gaps- SC}&:& \Delta_3\neq 0, \phi_1=\phi_2=\pi\\
 \mathrm {TRSB-SC}&:& \Delta_3\neq 0, \phi_1=\phi, \phi_2=-\phi\\
\mathrm {2\, gaps - SC}&:& \Delta_3=0, \phi_1=\pi/2, \phi_2=-\pi/2
\eea
Let us start from the path labeled by
(a) in Fig.\ \ref{fig-pd}, see Fig.\ \ref{fig-l112}. Here we identify a mode L1 which softens at
the $T_{TRSB}$ and remains always below the gap threshold. Above
$T_{TRSB}$ L1 is an ordinary Leggett mode associated to the phase
fluctuations in the two hole bands. Indeed, in this state the pairing
in each hole band is provided by the interband coupling to the third
electron band. Thus, within the hole-bands sector the problem is
formally equivalent to a two-gaps superconductor with dominant
intraband pairing, and the Leggett mode is well defined. Below the
$T_{TRSB}$ the SC order parameter in the hole bands becomes complex,
so that the Leggett-like oscillation drives also amplitude
fluctuations both in the hole and electron bands. Observe that below
$T_{TRSB}$ a second low-energy mode appears, labeled L2 in Fig.\
\ref{fig-l112}, which is only slightly below the gap threshold. Indeed,
at $T>T_{TRSB}$ this mode coincides with pure amplitude fluctuations
in the two hole bands, and thus it appears right at the gap edge
$2\Delta_{1,2}\equiv 2\Delta_{min}$. However, as one moves at higher
$\l/\eta$ values or one makes the two hole pockets inequivalent this
mode approaches rapidly the gap edge, becoming then overdamped. 
On the other hand at the full symmetric
point $\l=\eta$ L1 is exactly degenerate with the L2 mode. Indeed, at the $\l=\eta$ point the three bands are
completely equivalent, and the L1 and L2 describe the same
oscillation: the gaps in two bands approach each other, inducing a
change of modulus of the third gap.\cite{stanev_prb12}

\begin{figure}[htb]
\includegraphics[scale=0.3, angle=0,clip=]{fig5a.eps}
\includegraphics[scale=0.3, angle=0,clip=]{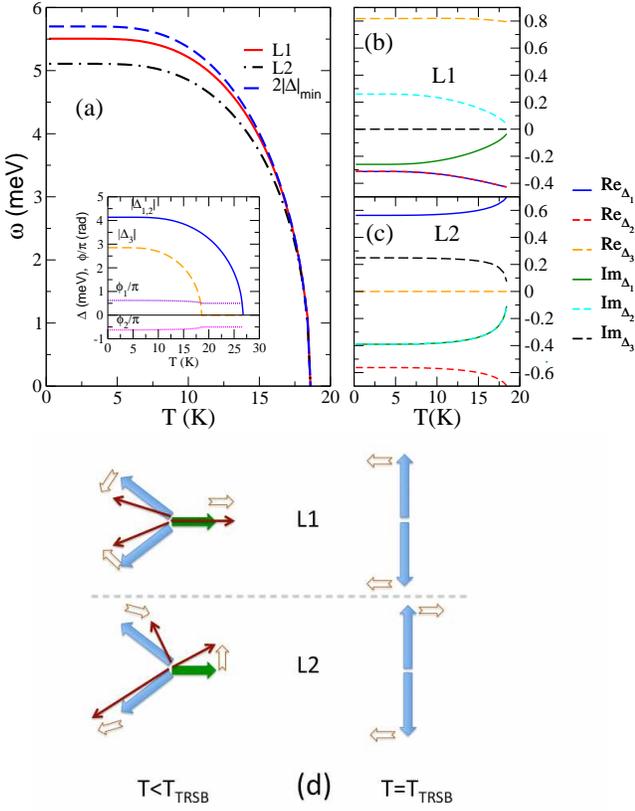}
\caption{(color online) Temperature evolution of the low-energy modes
  along the path (b) of Fig.\ \ref{fig-pd}, corresponding to
  $\l=0.92\eta$. Here $T_c=18.6$ K and $T_{TRSB}=23$ K. (a): Temperature
  dependence of the low-energy mode along with the minimum gap threshold,
  obtained by the temperature dependence of the gaps reported in the inset
  along with the phases of the hole gaps. (b) and (c): component of the
  eigenvectors corresponding to the modes labeled as L1 and L2 in panel
  (a). (d): schematic structure of the modes below and above
  $T_{TRSB}$, with the notation of Fig.\ \ref{fig-l112}. In contrast to the
  case $\l>\eta$ shown in Fig.\ \ref{fig-l112} here the fluctuations above
  $T_{TRSB}$ do not identify a mode. Nonetheless, exactly at $T=T_{TRSB}$
  L2 coincides with the Goldstone mode while L1 appears as an ordinary
  Leggett-like oscillation of the gaps in the two hole bands.}
\label{fig-l09}
\end{figure}

When one moves in the regime $\l<\eta$ (path (b) in Fig.\
\ref{fig-pd}) the role of the two modes in
the TRSB state changes and L2 becomes softer. More interestingly, at 
$T>T_{TRSB}$  the situation is completely
different in this case, since no soft mode can be found. This result
can be easily understood: at $T>T_{TRSB}$ the gap in the
electron band closes and the system is formally equivalent to a
two-band superconductor with dominant interband coupling. This is the
situation discussed in Sec. II, where no Leggett-like mode is
present since only one bonding SC channel exists. By close inspection
of the eigenvector components in Fig.\ \ref{fig-l09}b,c,d one sees that
as $T\ra T_{TRSB}^-$ the L2 mode tends to the Goldstone mode while the
L1 mode would coincide to the ordinary Leggett oscillation, which does
not identify a mode above $T_{TRSB}$ for the reason explained above. We then
recover the same result discussed below Eq.\ \pref{sphase3q0} in the
language of the $h_{1,2}$ fields, i.e. that exactly at $T=T_{TRSB}$ there are
two solutions at $\o=0$. However, while the Goldstone mode is always
well defined and it remains massless, the other solution can be
connected to a well-defined mode only below $T_{TRSB}$, where all the
three gaps are opened. In this respect, as soon as one modifies
slightly the coupling matrix \pref{defg3} in order to make the two
hole pockets inequivalent, the gap in the electronic band in 
general survive up to $T_c$. In this case, a soft mode can be found also in the
whole temperature interval $T_{TRSB}<T<T_{c}$, with  a similar
temperature dependence as the one shown in Fig.\ \ref{fig-l112}.

\begin{figure}[htb]
\includegraphics[scale=0.3, angle=0,clip=]{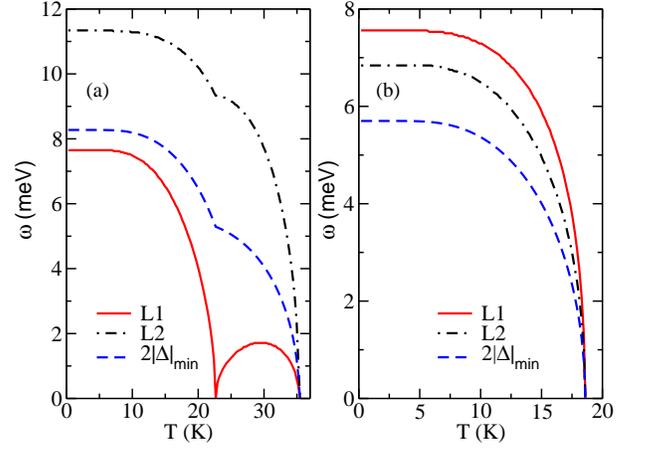}
\caption{(color online) Evaluation of the collective modes by means of
  a low-frequency expansion of the fermionic bubbles for the case (a)
  $\l=1.12\eta$ and (b) $\l=0.92\eta$, whose exact solutions are
  reported in Fig.\ \ref{fig-l112} and \ref{fig-l09},
  respectively. Notice that in both cases the frequencies of the
  low-energy modes are largely overestimated, and no soft mode is
  found for $\l=0.92$, in contrast to the correct result. }
\label{fig-small}
\end{figure}

It is worth noting that in evaluating numerically the collective modes
we retained the full frequency dependence of the electronic bubbles in
Eq.\ \pref{smatrix}. Indeed, the close proximity of one of the soft
modes in the TRSB phase to the gap edge makes the low-frequency
expansion\cite{stanev_prb12,maiti_prb13} dangerous, as observed also
in Ref.\ [\onlinecite{maiti_prb13}]. This is shown explicitly in Fig.\
\ref{fig-small}, where we report the temperature dependence of the
collective modes obtained by using the low-frequency approximation
\pref{l11}-\pref{l22} of the fermionic bubbles that appear in Eq.\
\pref{smatrix}. As one can see, while the massless character of the
$h_2$ fluctuations is correctly recovered at
$T=T_{TRSB}$, the absolute value of the low-energy
modes in the TRSB state is completely wrong in this approximation. In
particular, in the case $\eta<\lambda$ (Fig.\ \ref{fig-small}b) no
mode is found below the threshold for the quasiparticle
excitations. Even adding the next-order term in the low-frequency
expansion \pref{l11}, as suggested in Ref.\
[\onlinecite{maiti_prb13}], only one mode moves below the gap edge, 
in contrast to the correct result (Fig.\ \ref{fig-l09}a).

\begin{figure}[htb]
\includegraphics[scale=0.3, angle=0,clip=]{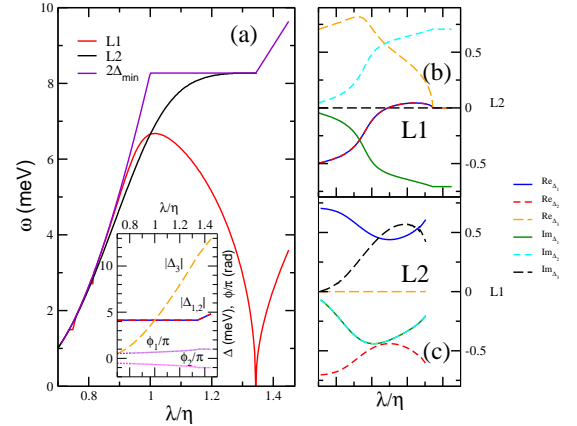}
\caption{(color online) (a) Evolution of the low-energy modes
  along the path (c) of Fig.\ \ref{fig-pd}, i.e. as a function of the ratio
  $\l/\eta$ at $T=0$. The corresponding gap and phase values are shown in
  the inset. (b) and (c): eigenvectors components of the two modes. Notice that at $\l=\eta$ the L1 and L2 mode are
  degenerate, as already observed
  before.\cite{stanev_prb12,maiti_prb13}. On the other hand as soon as one
  moves away from the symmetric point one mode moves rapidly towards the
  gap edge.}
\label{fig-x}
\end{figure}

To make also a closer connection to previous
work\cite{hu_prl12,stanev_prb12,maiti_prb13} we show in Fig.\ \ref{fig-x} the
evolution of the low-energy mode across the quantum TRSB transition,
i.e. the path labeled with (c) in Fig.\ \ref{fig-pd}. Here (see Fig.\
\ref{fig-x}a)  the
crossing between L1 and L2 at $\l=\eta$ is evident, and it is also
clear that in the regime $\l<\eta$ the vanishing of the electronic gap
will make it more difficult to resolve experimentally the soft
mode. Indeed, even if $\o_{L1}$ remains well below the gap in the hole
bands,  it rapidly approaches $2\Delta_{min}\equiv
2\D_3$. As we discussed above, this mode is missing in Ref.\
[\onlinecite{maiti_prb13}] since the authors used a low-energy expansion of the
fermionic bubbles in the numerical evaluation of the collective
modes. On the other hand, as soon as one makes
the hole pockets inequivalent\cite{maiti_prb13} the TRSB state
admits an end point at a finite critical value $\l_c^{min}$  also for
$\eta<\lambda$, so that all the
gaps are finite at $T_{TRSB}$ and only one mode becomes massless at
$\l_c^{min}$.  

\subsection{Experimental signatures of the Leggett mode}

Let us discuss now the relevance of the present results to
the experimental investigation of a TRSB state in pnictides. A 
natural probe for the identification of low-energy phase mode is
Raman spectroscopy, in full analogy with the case of the intra-band
dominated superconductor
MgB$_2$.\cite{blumberg_prl07,klein_prb10} Even
though a full calculation of the Raman response is beyond the scope of
the present manuscript,  by following the results of several
previous works,\cite{klein_prb84,klein_prb10,wu_prl09,hu_prl12} we 
outline the basic mechanism  which can make phase modes visible in Raman. 
Raman scattering allows one to measure
the response function for a charge density
$\tilde\rho(\bq)=\sum_\bk \gamma_\bk \rho(\bq)$ weighted with a
structure factor $\gamma_\bk$ that accounts for the specific geometry
of the incoming/outgoing light polarization. Since density and phase
fluctuations are conjugate variables,\cite{randeria_prb00,benfatto_prb04,
  sharapov_prb01,hu_prl12} phase fluctuations couple to the
Raman response as well, with some caveats on the allowed symmetry for
multiband superconductors.\cite{klein_prb10} In practice, this means that on top of the
bare Raman response due to quasiparticle excitations, that vanishes
below $2|\Delta_{min}|$ at $T\ll T_c$, collective phase excitations
manifest themselves in the Raman response as a peak at the typical
frequency $\o_L$ of the corresponding
modes.\cite{klein_prb10,wu_prl09,hu_prl12} More specifically, when
$\o_L$ lies below the treshold $2|\Delta_{min}|$ for quasiparticle
excitations the mode at low $T$ is weakly damped by residual
impurity-induced scattering processes, so the peak is sharp. In the
specific case of pnictides we have shown that Fig.\ \ref{fig-l112}
represents the typical thermal evolution of the phase modes across the
transition from the TRS to the TRSB state, see path (a) of Fig.\
\ref{fig-pd} . Since the Leggett-like mode which becomes massless at
the $T_{TRSB}$ lies always below $2|\D_{min}|$ it should be visible
already when entering the normal SC state, with a non-monotonic
temperature dependence: it first softens until $T_{TRSB}$ is reached
and then hardens again, saturating at low $T$, see $\o_{L1}(T)$ in
Fig.\ \ref{fig-l112}.  Observe that in Fig.\ \ref{fig-l112} the L1 mode
involves above $T_c$ fluctuations in the two hole bands, which are
assumed here for simplicity to have the same DOS, so that the
interaction anisotropy can be tuned by a simple parameter
$\l/\eta$. However, in real materials the two hole bands have
different DOS, and then different weighting factors $\gamma^l_\bk$,
which are simply proportional to the band masses.  According to
the discussion of Ref.\ [\onlinecite{klein_prb10}], this guarantess
that the L1 mode will be visible in Raman, despite it involves phase
fluctuations between bands having the same character.

We also verified that even in the
region at $\l>\l_{cr}\simeq 1.3$ of the phase diagram of Fig.\
\ref{fig-pd}, where the TRSB transition does not explicitly occurr,
the Leggett mode lies well below the gap treshold $2\D_{min}$ in a
wide range of temperatures and couplings, as shown also by the $T=0$ results reported
in Fig.\ \ref{fig-x}. Thus,  if in realistic materials with more
anisotropic interactions the phase
space where a TRSB state is realized will shrink,\cite{maiti_prb13}
making more difficult to realize a sample which displays the $s+is$
state, its proximity can be still evidenced by the emergence of a
phase mode at low (but finite) energy. In this situation, one can also
investigate the effects of the Leggett mode on other quantities, as
for example the
superfluid density. Indeed, by expanding the phase-only action
\pref{sphase3q0} at low frequency one finds that the Goldstone mode
and the Leggett one are coupled at finite frequency (see e.g. Eq.\
\pref{sphasesmall}). Thus, whenever the Leggett modes is massive it couples to
fluctuations of the overall SC phase, which in turn can affect the
temperature   dependence of the superfluid
density,\cite{benfatto_prb01,benfatto_prb04}   that has been recently
shown to be highly non-monotonic when 122 systems are strongly hole
doped.\cite{matsuda_cm13} This issue, which requires to account
properly also for density fluctuations and long-range Coulomb
interactions,\cite{benfatto_prb04} not included so far, will be the
subject of a future investigation.

\section{Discussion and Conclusions}

In the present manuscript we analyzed the behavior of
collective phase and amplitude modes in a multiband superconductor
with predominant interband pairing, which is the case physically
relevant for pnictide superconductors. The interband nature of the
pairing mixes bonding (attractive) and antibonding (repulsive) SC channels, that must be treated
with care while deriving the effective action for the SC fluctuations
by means of the standard Hubbard-Stratonovich decoupling. Here we
implement a generalized transformation $T$ of the multiband pairing
operators which has two crucial consequences:

(1) It allows to put the mean-field equations in a diagonal form:
\be
(\hat g^{-1}-\hat \Pi)\vec \D=0 \Ra \quad (\hat
\L^{-1} -T\hat \Pi T^T)\vec {\bar h}\equiv \sum_l Y_l \bar h_l =0
\ee
The saddle-point values of the HS fields $\vec {\bar h}$ characterize
the SC state. In the ordinary SC phase the eigenvalue $Y_1=0$
connected to the largest SC bonding channel  vanishes, and the
saddle-point value $\bar h_1$ of the corresponding HS
field is finite and can be taken real. The eigenvalue connected to the antibonding SC channel is
always positive, so that the corresponding HS field is always zero. In
this way, one sees that antibonding channels do not contribute to the SC
state. A
TRSB phase in the three-band model occurs when a second bonding SC
channel becomes active, so that the corresponding eigenvalue,
say $Y_2$, vanishes. In this situation the corresponding imaginary
part of the HS fields
$\bar h_2$
acquires a finite saddle-point value, and the gaps are
intrinsically complex:
\bea
Y_1=0 &\Ra& \mathrm{TRS-SC}, \bar h_1=R_1\\
Y_1=Y_2=0 &\Ra& \mathrm{TRSB-SC}, \bar h_1=R_1, \bar h_2=iI_2
\eea

(2) The low-energy behavior of the collective phase fluctuations is
  uniquely determined by the mean-field equations. More specifically,
  one sees that at $q=0$ the fluctuations in the TRS phase sectors are
  described as
\be
S^P_{FL} (q=0)\sim \sum_l Y_l (Im h_l)^2
\ee
This has several implications: (i) the fluctuations of the ordering
field $Im h_1$ are trivially massless  ($Y_1=0$) since this is the Goldstone mode of the SC
transition; (ii) for the two-band case the fluctuations
on the antibonding HS fields  do not identify any collective mode, so
that for interband-dominate pairing the Leggett mode is absent, in contrast
to ordinary intraband-dominated superconductors as MgB$_2$; (iii)
in the three-band case the transition to a TRSB phase ($Y_2=0$) is uniquely
associated to {\em massless} phase fluctuations described by the
imaginary part of the ordering field
$h_2$. 

The absence of the Leggett mode in a two-band model for pnictides 
 can be understood on physical grounds by having in mind
the marked difference between a two-band superconductor with dominant
intra- or inter-band pairing. In the former case one has two bonding
SC channels: the SC transition is controlled as usual by the one giving rise to a
larger condensation energy, i.e. larger gaps. On the other hand, there
exists a second possible solution with smaller gap values: the Leggett mode
describes indeed deviations from the ground state in the direction of this
second possible solution. For example, when the interband pairing is
positive the two gaps $\D_1,\D_2$ have the same sign, i.e. the same
saddle-point phases $\bar\vartheta_1=\bar\vartheta_2$. In this situation, the
Leggett mode identify phase fluctuations of opposite sign in the two bands, that
would lead the system towards the solution $\bar\vartheta_1=-\bar\vartheta_2$ with
higher energy, as reflected in the massive character of the Leggett
mode. In this respect, antiphase oscillations can be put in resonance with
the additional SC channel, so that they identify a collective mode. On the other hand, when the coupling is predominantly
interband there is no additional SC channel in the problem, so that the
same kind of oscillations do not identify any proper mode of the system.

For the three-band case, it must be emphasized that the massless character of the Legget phase mode
found in item (2) above holds regardless the
specific nature of the two SC bonding channels which become degenerate
at the TRSB transition. Thus, it can be applied as well also to other
(multiband or multichannel)
systems where a TRSB phase as been predicted, as e.g. highly doped
graphene\cite{chubukov_natphys12,thomale_graphene_prb12,lee_prb12},
water-intercaled sodium cobaltates\cite{thomale_cm13}  
and locally noncentrosymmetric SrPtAs.\cite{sigrist_prb12} The last 
system is particularly promising in this respect, since recent
muon spin rotation ($\mu SR$) measurements could be actually interpreted as evidence of a
TRSB phase.\cite{biswas_prb13} Indeed, the massless character of the phase mode at the
transition is a unique signature of the TRSB transition, that can be
used to rule out other possible interpretation of the $\mu SR$ data. 

In the specific case of pnictides, by performing an explicit numerical
calculation of the collective modes we also established some
additional results. In particular, we found that  below $T_{TRSB}$ the massless mode becomes massive again,
while a second low-energy mode can appear in some range of
parameters. For the case of pnictides, by scanning in temperature a given
sample which is either in case (a) or (b) of Fig.\ \pref{fig-pd} the
appearance of a mode below the gap threshold will signal the emergence
of the TRSB state. Since phase fluctuations couple to
density
fluctuations,\cite{legget,benfatto_prb04,klein_prb84,klein_prb10,hu_prl12}
the most direct probe of this mode is via 
Raman scattering, in analogy with the result found for
MgB$_2$.\cite{blumberg_prl07}. Even
though a full calculation of the Raman response is beyond the scope of
the present manuscript, we expect that in the realistic case of
pnictides the temperature evolution of the Raman
spectra can offer a powerful mean to the identification of a TRSB
state, shedding then new light on the
theoretical understanding of the pairing mechanism itself in these
unconventional superconductors.

\section{Acknowledgements}
We thank S. Caprara for useful discssions and suggestions. 
The authors acknowledge financial support  by the Italian
MIUR  under the project FIRB-HybridNanoDev-RBFR1236VV and by the
Spanish Ministerio de Economía y Competitividad (MINECO) under the
project FIS2011-29680.

\appendix
\section{Equivalence between the Gaussian action derived in cartesian or
  polar coordinates}
In this Appendix we show that in the single-band case the Gaussian action for amplitude and phase
fluctuations can be equally derived by using a cartesian (real and
imaginary part) or polar (amplitude and phase) description of the SC
fluctuations. In the multiband case, as we discussed in the present
manuscript, one chooses as convenient variables (i.e., as HS fields) proper combinations of
the gaps in the various bands. Nonetheless, when all the the SC
channels are bonding one can still introduce a set of HS fields associated to
the fluctuations in each band. Indeed, in this case the equivalent of
the relations \pref{dreal}-\pref{dimag} between the physical fields
and the HS fields do not contain imaginary units. Thus, they are not
only valid for the average values, but they  represent a change of variables for bosonic fields in the
functional integral defining the partition function. This case will be
discussed indeed in Appendix B, where we explicitly apply our
formalism to the two-band case with predominant intraband pairing. However, as soon
as at least one channel is antibonding the only possible route is to use
collective HS fields properly defined in each channel, as we have done
in the present manuscript.  Still, the general relations between the
fermionic bubbles that we will establish below can be used to
interpret the low-energy behavior of the collective modes. In
particular, it will be useful to compare the 
derivation of collective modes in the TRSB phase given in Refs. [\onlinecite{ota_prb11,hu_prl12}],
based on a polar description, and the one presented here, based on the use of 
cartesian coordinates, equivalent also to the approach of Refs.\
[\onlinecite{stanev_prb12,maiti_prb13}].

Let us start again from Eq.\ \pref{ham2} written for a single-band
superconductor, and let us decouple the interaction term by means of the HS
decoupling \pref{hs}, by introducing an HS field $\Delta(x)$. The equivalent of Eq.\ \pref{snew} will then read:
\be
\lb{shs}
S=S_0+\int d\tau d\bx \frac{|\Delta(x)|^2}{g}-\left( \Delta^*c_\down
  c_\up+h.c.\right)
\ee
After integration of the fermions the saddle-point value $\D$ ( chosen
to be real) of the HS field will appear in the mean-field Green's function
$\bar G$, while its fluctuations will be decomposed as $\Delta_\bq=Re
\Delta_\bq+iIm\Delta_\bq$, so that Eqs.\ \pref{defgbar} and \pref{defsigma}
will be explicitly given by:
\begin{eqnarray}
\lb{defgbar1}
\bar{G}_{k}^{-1} & = & \begin{pmatrix}i\omega_{n}-\xi_{k} &  \D\\
 \Delta & i\omega_{n}+\xi_{k}
\end{pmatrix},\\
\lb{defsigma1}
\Sigma_{q=k-k'} & = & \sqrt{\frac{T}{V}} \left(Re \Delta_q\sigma_1+Im \Delta_q\s_2\right),
\end{eqnarray}
where $\D$ is the solution of the self-consistency equation:
\be
\lb{selfcons}
\Pi=\frac{1}{g}\Ra \frac{1}{V}\sum_\bk \tanh \frac{\beta E_\bk}{2}
\frac{1}{2E_\bk}=\frac{1}{g},
\ee
where $\Pi$ is the Cooper bubble \pref{defcoop} and
$E_\bk=\sqrt{\xi_\bk^2+\D^2}$. 
The explicit connection established by Eq.\ \pref{defsigma1}  above between the real and imaginary part of the
fluctuating field and the $\sigma_1,\sigma_2$ Pauli matrices, respectively,
allows one to derive immediately the coefficients of the 
action $S_{FL}$ \pref{sfl} at Gaussian order:
\be
\lb{sflc}
S_{FL}^{Car}=\frac{1}{2}\sum_q \eta^T_{-q} 
\begin{pmatrix}  \L^{11}(q)+\frac{2}{g} &  \L^{21}(q)\\
\L^{12}(q) &  \L^{22}(q)+\frac{2}{g} 
\end{pmatrix}\eta_q,
\ee
where in analogy with Eq.\ \pref{sflg} we defined $\eta^T_q=(Re\D_q,
Im\D_q)$. The fermionic bubbles are defined as:
\be
\lb{deflij}
\L^{ij}(q)=\frac{T}{V}\sum_k \mathrm{Tr} \left[\bar G_{k+q}\s^i\bar G_{k}\sigma^j\right].
\ee
More specifically we have:
\begin{widetext}
\bea
\L^{ij}(q)&=&\frac{1}{V}\sum_\bk
\left[\frac{(uu)_{ij}}{E_\bk'-E_\bk-i\O_m}+\frac{(vv)_{ij}}{E_\bk'-E_\bk+i\O_m}
\right][f(E_\bk')-f(E_\bk)]+\nn\\
\lb{gen}
&+&\left[\frac{(uv)_{ij}}{E_\bk'+E_\bk-i\O_m}+\frac{(vu)_{ij}}{E_\bk'+E_\bk+i\O_m}
\right][f(E_\bk')-f(-E_\bk)]
\eea
\end{widetext}
where $E_\bk'=E_{\bk+\bq}$ and the coherence factors are given by:\cite{kulik_jltp81}
\bea
(uu)_{11}=(vv)_{11}&=&\frac{1}{2}\left(
  1-\frac{\xi_\bk'\xi_\bk-\D^2}{E_\bk'E_\bk}\right)\\
(uv)_{11}=(vu)_{11}&=&\frac{1}{2}\left(
  1+\frac{\xi_\bk'\xi_\bk-\D^2}{E_\bk'E_\bk}\right)
\eea
\bea
\lb{u22}
(uu)_{22}=(vv)_{22}&=&\frac{1}{2}\left(
  1-\frac{\xi_\bk'\xi_\bk+\D^2}{E_\bk'E_\bk}\right)\\
\lb{v22}
(uv)_{22}=(vu)_{22}&=&\frac{1}{2}\left(
  1+\frac{\xi_\bk'\xi_\bk+\D^2}{E_\bk'E_\bk}\right)\\
(uu)_{12}=-(vv)_{12}&=&\frac{1}{2}\left(\frac{\xi_\bk}{E_\bk}-\frac{\xi_\bk'}{E_\bk'}\right)\\
(uv)_{12}=-(vu)_{12}&=&-\frac{1}{2}\left(\frac{\xi_\bk'}{E_\bk'}+\frac{\xi_\bk'}{E_\bk'}\right)
\eea
As one can easily check, $\L^{12}(q)\sim \O_m{\cal O}(q^2)$, so that in the static limit it can be
neglected,  leading to the effective decoupling
between the amplitude and phase fluctuations that has been used in
Sec. IIB. Moreover, from Eq.\ \pref{gen} and \pref{u22}-\pref{v22}
above it follows that:
\be
\L^{22}(0)=-\frac{1}{V}\sum_\bk \tanh \frac{\beta E_\bk}{2}
\frac{1}{E_\bk}=-2\Pi=-\frac{2}{g},
\ee
where we used the self-consistency equation \pref{selfcons}
above. Thus, one immediately recovers the massless character of the $Im \D_q$
fluctuations from Eq.\ \pref{sflc}. Notice that since we did not
introduce explicitly the density fluctuations, other phase modes like
the Carlson-Goldman one\cite{carlson_prl75,sharapov_prb02,efetov_prb07} cannot be
explicitly obtained. On the other hand, these sound-like modes are
usually relevant only at high temperature or strong disorder,\cite{efetov_prb07} unless
the gap has nodes\cite{sharapov_prb02}, that are not the cases
relevant for the present discussion.

Let us now discuss the derivation of the Gaussian fluctuations within the
polar-coordinate scheme.\cite{benfatto_prb04} In this case, before
integrating our the fermions in Eq.\ \pref{shs}, one can make explicit the dependence on the phase
$\theta$ of the HS field by means of a Gauge transformation on the
fermionic fields, $c_\s(x)\ra c_\s(x)e^{i\theta/2}$. As a consequence, while
$\bar G_k$ is unchanged, the self-energy $\Sigma_{kk'}$ describing the SC
fluctuations will be expressed in terms of the variables
$|\Delta|_q,\theta_q$:\cite{benfatto_prb04}
\bea
\Sigma_{kk'}&=&\frac{T}{V}|\D|_{k-k'}\s_1+
 \sqrt{\frac{T}{V}}\frac{i}{4m}(\bk-\bk')\cdot(\bk+\bk')\theta_{k-k'}\s_0+\nn\\
&+&\sqrt{\frac{T}{V}}\frac{\o_{k-k'}}{2}\theta_{k-k'}\s_3+\nn\\
\lb{sigmap}
&+&\frac{T}{V}\sum_\bs
\frac{(\bk-\bs)\cdot(\bs-\bk')}{8m}\theta_{k-s}\theta_{s-k'} \s_3
\eea
As one can see, in this case the self-energy contains only spatial and
time derivatives of the phase, associated respectively to the Pauli
matrices $\sim \bk \s_0$ and $\s_3$, which describe in Nambu formalism the
fermionic current and density.  Thus, using Eq.\ \pref{sigmap} in the
expansion \pref{sfl} one can write the effective Gaussian action
as:
\be
\lb{sflp}
S_{FL}^{Pol}=\frac{1}{2}\sum_q \zeta^T_{-q} 
\begin{pmatrix}  \L^{11}(q)+\frac{2}{g} & -\frac{i}{2}q_\mu \L_{J1}^\mu \\
 -\frac{i}{2}q_\mu \L_{1J}^\mu &  \frac{1}{4} q_\mu q_\n \tilde \L^{\mu\nu}_{JJ}(q) 
\end{pmatrix}\zeta_q,
\ee
where $\zeta_q\equiv (|\D|_q,\theta_q)$. Here we used the quadrivector notation $q^\mu=(i\O_m,\bq)$,
$q_\mu=(i\O_m,-\bq)$ and we  introduced the generalized current-current bubbles

\bea
\lb{defkij}
\tilde \L_{JJ}^{\m\n}&=&-\frac{n}{m}\eta^{\m\n}(1-\eta^{\n 0})+\L_{JJ}^{\m\n},\\
\lb{defcij}
\L_{JJ}^{\m\n}&=&\frac{T}{V}\sum_k \mathrm{Tr} \left[\bar
G_{k+q}\g^\mu(k,k+q)\bar G_k \g^\n(k+q,k)\right],\\
\lb{defji}
\L^\mu_{Ji}&=&\frac{T}{V}\sum_k \mathrm{Tr} \left[\bar
G_{k+q}\g^\mu(k,k+q)\bar G_k \s_i\right],
\eea
where $\eta^{\m\n}=\mathrm {diag}(1,-1,-1)$ and the current vertex is 
\be
\lb{vertex}
\g^\mu(k,k+q)=\left(\s_3, \frac{\bk+\bq/2}{m} \s_0\right).
\ee
Notice that the above definitions \pref{deflij} and \pref{defcij}-\pref{defji} are slightly redundant, since
e.g. $\L^{00}_{JJ}\equiv \L_{33}$. Nevertheless, Eqs.\
\pref{defcij}-\pref{defji} allow for a transparent
interpretation of the phase mode in Eq.\ \pref{sflp}, whose dispersion 
is given explicitly by:
\be
\lb{prop}
q_\mu q_\n \tilde \L^{\mu\nu}_{JJ}(q)\equiv -\O_m^2 \tilde
\L^{00}_{JJ}+q^iq^j\tilde \L_{JJ}^{ij}-2i\O_mq^i\tilde\L_{JJ}^{0i}.
\ee
Indeed, 
since only time or spatial derivatives of the phase field enter
the self-energy \pref{sigmap}, the hydrodynamic limit of the phase
mode is easily obtained by taking the
$q=0$ limit of the current-current fermionic bubbles which appear as
coefficients in Eq.\ \pref{prop}. More specifically, since $\L_{JJ}^{00}(0)=-\kappa$,
where $\kappa$ is the compressibility,
$\L_{JJ}^{ij}(0)=(\rho_s/m)\d_{ij}$ and $\L_{JJ}^{0i}(0)=0$, one finds
immediately that in the long-wavelength limit the
 phase mode has the well-known sound-like dispersion
\be
\lb{BA}
q_\mu q_\n \tilde \L^{\mu\nu}_{JJ}(q)\simeq  -\o^2\kappa+\bq^2
\frac{\rho_s}{m}
\ee
which characterizes the Bogoliubov-Anderson
  mode. Instead, in the cartesian notation of Eq.\ \pref{sflc} the
  dispersion of the phase mode must be obtained by performing the
  low-$q$  expansion of the $\L_{22}(q)$ bubble. 
An alternative root is to exploit the equivalence
  between the two derivations \pref{sflc} and
\pref{sflp}, that must be valid at all orders in $q$. Here we prove
explicitly this
equivalence by using the definitions of the fermionic bubbles and the
identity\cite{chien_cm12}: 
\be
\lb{equiv}
\bar G^{-1}_{k+q}\s_3-\s_3 \bar G^{-1}_{k}=q_\mu\g^\mu(k+q,k)-2i\bar\Delta
\s_2.
\ee
One can then prove the three equalities:
\bea
\lb{eq1}
q_\mu \L^{\mu}_{J2}(q)&=&2i\bar \Delta \left[
  \L_{22}(q)+\frac{2}{g}\right],\\
\lb{eq2}
q_\mu \tilde\L_{JJ}^{\mu\nu}(q)&=&2i\bar\Delta \L^\nu_{2J}(q),\\
\lb{eq3}
q_\mu \tilde\L_{J1}^{\mu}&=&2i\bar\Delta \L_{21}.
\eea
Let us show for example the demonstration of Eq.\ \pref{eq1}. By means of
the definitions \pref{deflij} and \pref{defji} and the equivalence \pref{equiv} we have:
\bea
& & q_\mu \L^{\mu}_{J2}(q)-2i\bar \Delta \L_{22}=\nn\\
&=&\frac{T}{V}\sum_k \mathrm{Tr} \left[\bar
G_{k+q}(q_\mu\g^\mu(k,k+q)-2i\bar\Delta\s_2)\bar G_k \s_2\right]=\nn\\
&=&\frac{T}{V}\sum_k \mathrm{Tr} \left[\s_3\bar
G_{k}\s_2\right]-\frac{T}{V}\sum_k \mathrm{Tr} \left[\bar
G_{k+q}\s_3\s_2\right]=\nn\\
&=&2i \frac{T}{V}\sum_k \mathrm{Tr} \left[\bar G_k \s_1\right]=4i\bar\Delta \Pi=\frac{4i\bar\Delta}{g}
\eea
where we used the definition of the $\Pi$ bubble, $\Pi\equiv
\frac{T}{2V}\sum_k \mathrm{Tr} \left[\bar G_k \s_1\right]$ and the
self-consistency equation \pref{selfcons} $\Pi=1/g$. The remaining equalities
\pref{eq2}-\pref{eq3} can be obtained with a similar
procedure.\cite{chien_cm12} By means of Eq.s\ \pref{eq1}-\pref{eq2},
and using $\L_{2J}^\mu(q)=\L_{J2}^\mu(-q)=-\L_{J2}^\mu$ we then have:
\be
\lb{l22equiv}
\frac{1}{4}q_\mu q_\n \tilde \L^{\mu\nu}_{JJ}(q)=-\frac{i}{2}\bar \D
q_\nu\L^\nu_{J2}(q)=\bar\D^2 \left[
  \L_{22}(q)+\frac{2}{g}\right]
\ee
so that, since $Im \D_q=\bar\Delta\theta_q$, we recover the
equivalence between the phase-fluctuation propagator in Eqs.\
\pref{sflc} and \pref{sflp}. Analogously, by means of Eq.\ \pref{eq3}
we recover the equivalence in Eq.s\ \pref{sflc} and \pref{sflp} between the off-diagonal terms in the amplitude-phase
fluctuations, completing the demonstration of the full equivalence of
the two procedures. Notice that the above relations \pref{BA} and
\pref{l22equiv} allow one to easily derive the low-momentum expansion of
the $\L_{22}$ bubble, that has been used in Eq.\ \pref{l22} of the main
text. Indeed, we have that:
\bea
\L^{22}(q)&=& -2\Pi+\frac{1}{4\D^2}q_\mu q_\n \tilde
\L^{\mu\nu}_{JJ}(q)\simeq\nn\\
\lb{l22lowq}
&=&-2\Pi+\frac{1}{4\D^2}\left(\O_m^2\kappa +\bq^2 \frac{\rho_s}{m}\right)
\eea
Notice that in Eq.\ \pref{l22lowq} above the coefficient of the $\O_m^2$
term, i.e. the density-density correlation function $\L_{JJ}^{00}(q)$, has
been taken in the static limit, where it gives the compressibility. On the
other hand, as it is well known,\cite{sharapov_prb01}
 at finite $T$ the dynamic limit of $\L_{JJ}^{00}(q)$, which is the one relevant to
compute the collective modes at $\bq=0$ in Sec. II and III, differs from
$\kappa$. On the other hand, the low-$q$ expansion of the $\L^{22}(q)$  can
still be connected to the generalized current-current bubbles, that is the
relation needed to recover the equivalence between the two derivations of
the collective modes.

It is worth noting that in the single-band case the description in terms of cartesian or polar
coordinates is equivalent since one usually chooses a gauge where the
saddle-point gap value $\bar\D$ is real. Thus, the two descriptions correspond to the same choice
of fluctuations directions and are then trivially equivalent. On the
other hand, if one had chosen a finite saddle-point phase for the gap then the
cartesian description of fluctuations would be very inconvenient,
since in this case the $\L_{12}(q=0)$ bubble which couples real and
imaginary parts in Eq.\ \pref{sflc} would be non zero, as we already
emphasized in Eq.\ \pref{l12t} above for the TRSB. However, in the
single-band case one could still make a proper rotation to 
polar coordinates that would lead again to decoupled 
amplitude and phase fluctuations. The multiband TRSB case discussed in
Sec. IIIB is instead different. Indeed, in the TRSB state each gap
acquires a non-trivial saddle-point value for the phase. In this
situation, even thought the choice of polar coordinates
for each band could still make the part of fluctuations described by
fermionic bubbles ortogonal, the coupling matrix $\hat g_{lm}$ in Eq.\
\pref{smatrix} will not be diagonalized by this rotation, making amplitude and phase fluctuations always
intrinsically mixed.

Finally, let us derive for the sake of completeness the relations between
the average values of the HS field $\Delta_q$ and the physical correlation
functions expressed in terms of the fermionic operators $\phi_q$
defined in Eq.\ \pref{defffields} above.\cite{depalo_prb00} Let us discuss it
for the case of cartesian coordinates: starting from Eq.\ \pref{shs} we add
a source field $\Psi_\bq$ which couples to $\phi_q$ such that:
\bea
\lb{zcompl}
Z&=&\int\mathcal{D}c_{\sigma}\mathcal{D}c_{\sigma}^{{\dagger}}\mathcal{D}\D\mathcal{D}\D^{\dagger}\,
e^{-S+\sum_q (\phi^\dagger_q \Psi_q+h.c.)}\nn\\
\lb{phiave}
\langle \phi_q\rangle &=& \left.\frac{\pd \ln Z}{\pd \Psi^*_q}\right |_{\Psi=0},\\
\lb{phiphi}
\langle \phi^\dagger_q \phi_{-q}\rangle &=& \left. \frac{\pd^2 \ln Z}{\pd \Psi_q\Psi^*_{-q}}\right |_{\Psi=0}.
\eea
To perform explicitly the derivatives on the right-side of Eqs.\
\pref{phiave}-\pref{phiphi} we can notice that the total action in Eq.\
\pref{zcompl} can be written as:
\bea
S'&=&S-\sum_q (\phi^\dagger_q \Psi_q+h.c.)=\nn\\
&=&\sum_q \frac{|\D_q|^2}{g}-\sum_q
\left[ (\Psi_q^*+\D^*_q)\phi_q+h.c.\right]=\nn\\
&=&\sum_q \frac{|\tilde\D_q|^2}{g}-\sum_q
\left[ \tilde\D^*_q\phi_q+h.c.\right]+\nn\\
&+&\sum_q \frac{|\Psi_q|^2}{g}-\frac{(\tilde\Delta^*_q\Psi_q+h.c.)}{g},
\eea
where we shifted $\Psi_q+\Delta_q=\tilde\Delta_q$. One can then easily
derive the relations:
\bea
\left.\frac{\pd \ln Z}{\pd \Psi^*_q}\right |_{\Psi=0}&=&\frac{1}{g}\langle
\Delta_q \rangle=\langle \phi_q\rangle,\\
\left. \frac{\pd^2 \ln Z}{\pd \Psi_q\Psi^*_{-q}}\right |_{\Psi=0}&=&-\frac{1}{g}+\frac{1}{g^2}\langle
\Delta^*_q\Delta_{-q} \rangle=\langle \phi^\dagger_q \phi_{-q}\rangle,
\eea
where we also used the fact that at $\Psi=0$ the averages of the $\Delta_q$
and $\tilde\D_q$ fields coincide. By direct inspection on the Gaussian
action \pref{sflc} for the HS-field fluctuations, we recover the well-known
result that the correlator for the HS field $\Delta_q$ corresponds to the
RPA resummation of the potential, while the correlator of the physical
field gives the RPA resummation on the corresponding fermionic
susceptibility. For example for the real components we have:
\bea
\langle Re\D_{-q}Re\D_q \rangle&=&\frac{g}{1+\frac{g}{2}\L^{11}(q)},\\
\langle Re\phi_{-q}Re\phi_q \rangle&=&\frac{-\L^{11}(q)/2}{1+\frac{g}{2}\L^{11}(q)}.
\eea

\section{Derivation of the Leggett mode for dominant intraband pairing}
In this Appendix we show explicitly how the generalized transformation
\pref{grot} can be used to obtain the Leggett 
mode in the two-band case with dominant {\em intraband} pairing, as it would
be appropriate for example to MgB$_2$.\cite{sharapov_epjb02,efetov_prb07}. As it
has been shown in Ref.\ [\onlinecite{sharapov_epjb02}] in this case
one could easily obtain the Leggett's mode dispersion by using a
straightforward generalization to a two-band case of the derivation
\pref{sflp} reviewed above in terms of
polar coordinates. By introducing a HS field for each band, with
phases $(\theta_1(q),\theta_2(q))$, the long-wavelength $\bq=0$
phase fluctuations (decoupled from the amplitude ones) are described by the matrix:
\be
\hat S_{FL}^P=\frac{1}{8}\left(
\begin{array}{cc}
-N_1\o^2+A & -A\\
-A & -N_2\o^2+A
\end{array}
\right)
\lb{eqm}
\ee
where $A$ is a constant connected to the matrix $\hat g$ of the SC
couplings and to the saddle-point gap values $\bar \D_l$ in each band
\be
A=\frac{8g_{12}\bar\D_1\bar\D_2}{det \hat g}.
\ee
In Eq.\ \pref{eqm} we recognize the $\o^2$ expansion derived in Eq.\
\pref{prop} above, with the compressibility $\kappa_l$ in each band
approximated by the corresponding density of states $N_l$. The phase
collective modes are found as usual as solutions of the equation $det 
\hat S_{FL}^P=0$. The first solution $\o=0$ corresponds as to the
Bogoliubov-Anderson (BA) 
mode, while a second solution exists corresponding to the Leggett mode\cite{legget}:
\be
\lb{omll}
\o^2=\o^2_L=A\frac{N_1+N_2}{N_1N_2}.
\ee
Notice that, as already observed by Leggett in his original
paper,\cite{legget}  this solution only exists when $A>0$, i.e. when $det\hat
g>0$ (since $sign(\bar \D_1\bar \D_2)=sign \, g_{12}$), which
corresponds to {\em intraband}-dominated coupling.   One can also
easily verify that the BA mode corresponds to fluctuations having
$\theta_1=\theta_2$, while the Leggett mode corresponds to
\be
(-N_1\o^2_L+A)\theta_1-A\theta_2=0 \Ra \frac{\theta_1}{\theta_2}=-\frac{N_2}{N_1}
\lb{legget}
\ee
i.e. to antiphase oscillations in the two bands, weighted with the
respective DOS. As usual, the Leggett mode (as well as the
Carlson-Goldman one)  can be equally found in
linear-response theory.\cite{efetov_prb07}

Let us rewrite instead the effective phase-only action after using the
generalized transformation $T$. In this case, since both the
eigenvalues $\L_1>\L_2>0$ of $\hat g$ are positive, the hyperbolic matrix in Eq.\
\pref{defhp} will be replaced by an ordinary rotation matrix, which
preserves the structure $diag(1,1)$ present in this case. The $T$ matrix
will again be used to decouple the mean-field equations for the HS fields:
\be
\lb{sss}
{\small 
\begin{pmatrix}\frac{1}{\Lambda_{1}}-\sum_lT_{1l}^{2}\Pi_{l} & 0 \\
0 & \frac{1}{\Lambda_{2}} -\sum_lT_{2l}^{2}\Pi_{l}
\end{pmatrix}
\begin{pmatrix} \bar h_1 \\\bar h_2
\end{pmatrix}}=Y_1 \bar h_1+Y_2\bar h_2=0.
\ee
where now no imaginary unit is associated to $\bar h_2$ since it decouples
an ordinary bonding SC channel, and also $Y_2$ could in principle
vanish. However, as soon as $Y_1=0$ and one enters the SC state $Y_2$ can
never vanish (see Appendix C), so the $\bar h_1$ will still be the only
order parameter of the SC transition. 
The
effective action will have the same structure of Eq.\
\pref{sphase} derived above, with the remarkable difference that now
the HF field $h_2$ used to decouple the $\L_2$ channel will not carry out
an additional $i$ unit, so that the $-1/\L_2$ term in Eq.\ \pref{sphase} is
replaced by $+1/\L_2$:
\bea
\hat S^P_{FL}(q)=
\begin{pmatrix}
\frac{1}{2}\sum_{l}\Lambda_{l}^{22}(q)T_{1l}^{2}+\frac{1}{\Lambda_{1}}
& 
 \frac{1}{2}\sum_{l}\Lambda_{l}^{22}T_{1l}T_{2l}\\
 \frac{1}{2}\sum_{l}\Lambda_{l}^{22}T_{1l}T_{2l} &
\frac{1}{2}\sum_{l}\Lambda_{l}^{22}(q)T_{2l}^{2}+\frac{1}{\Lambda_{2}}
\end{pmatrix}\nn\\
\lb{sphasel}
\eea
To compute the above Eq.\ \pref{sphasel} at $\bq=0$ and small $\o$ we
use the expansion \pref{l22} of the $\L_{22}$ bubbles, so that we obtain:
\be
\lb{sflpt}
\hat S_{FL}^P=-\left(
\begin{array}{cc}
B\o^2  & C\o^2\\
C\o^2 & D\o^2-m^2 
\end{array}
\right)
\ee
where
\bea
\lb{b}
B&=&\frac{1}{8}\sum_l \frac{N_l}{\D_l^2}T^2_{1l}\\
\lb{d}
D&=&\frac{1}{8}\sum_l\frac{N_l}{\D_l^2}T^2_{2l}\\
\lb{c}
C&=&\frac{1}{8}\sum_l\frac{N_l}{\D_l^2}T_{1l}T_{2l}\\
\lb{m2}
m^2&=&\frac{1}{\L_2}-\sum_l\Pi_lT_{2l}^2
\eea
Notice that $m^2$ in Eq.\ \pref{m2} above in nothing else than the second
eigenvalue $Y_2$ of the matrix of mean-field equations \pref{sss} above,
which is always non-zero below $T_c$ for a system with finite interband
coupling.  
It is then clear that also the matrix \pref{sflpt} leads to two
solutions. The first one at   $\o=0$ corresponds to the BA mode: it
involves only fluctuations of the $Im h_{1,q}$ field, which thanks to
the gap definitions \pref{mfgap} is indeed an uniform phase rotation
for the gaps in both bands, see Eq.\ \pref{defBA} above. The second solution is found at the
frequency:
\be 
\lb{oml}
\o^2=\frac{Bm^2}{BD-C^2}.
\ee 
We will now show that Eq.\ \pref{oml} coincides with the expression
\pref{omll} above, by deriving the explicit expressions of the $T$
matrix from the three conditions established in Sec. II: $det \, T=1$,
Eq.\ \pref{fix} $\sum_l T_{1l}\Pi_lT_{2l}=0$ and the saddle-point
equation \pref{saddle}, which defines also the mean-field gaps
$\D_l$ in Eq.\ \pref{mfgap}. One can then easily show that 
\be
\lb{tmatrix}
T=\frac{1}{\sqrt{\sum_l \Pi_l\D_l^2}}\begin{pmatrix} 
\D_1/\sqrt{\L_1} & \D_2/\sqrt{\L_1}\\
-\sqrt{\L_1}\Pi_2\D_2 & \sqrt{\L_1}\Pi_1\D_1
\end{pmatrix}, 
\ee
where we also have that the saddle-point value of the HS ordering
field $\bar h_1$ is given by
\be
\lb{h1bar}
\bar h_1=\L_1 
\sum_l \Pi_l\D_l^2
\ee
By means of Eqs.\ \pref{tmatrix}-\pref{h1bar} we can then express the
coefficents \pref{b}-\pref{m2} in terms of the gap values $\D_l$ and of
the eigenvalues, connected to the matrix $\hat g$ of the SC
couplings. With lengthly but straightforward calculations we then
have:
\bea
B&=&\frac{1}{8\bar h_1^2} \sum_l N_l,\\
D&=&\frac{\L_1^2}{8}
\frac{N_1\Pi_2^2\D_2^4+N_2\Pi_1^2\D_1^4}{\D_1^2\D_2^2},\\
C&=&\frac{\L_1}{8\bar h_1} \frac{-N_1\Pi_2\D_2^2+N_2\Pi_1\D_1^2}{\D_1\D_2},\\
m^2&=&\frac{\L_1}{\bar h_1}\left(1-\Pi_1\Pi_2 \mathrm {det} \hat
  g\right)=\frac{g_{12}}{\mathrm{det} \hat g \D_1\D_2}.
\eea
As a consequence, we get in Eq.\ \pref{oml} that
$(BD-C^2)/B=N_1N_2/8\D_1^2\D_2^2(N_1+N_2)$ and we then recover the
expression \pref{omll} for the frequency of the second eigenmode. By means of the same relations one can also
prove that the eigenvector corresponding to the \pref{oml} solution,
i.e. $Im h_1=-(C/B) Im h_2$ describes the antiphase fluctuations
\pref{legget} identified above for the Leggett mode. 

\section{TRSB transition in a three-band model}
In this Appendix we discuss the TRSB transition in the three-band case in
terms of the action for the HS fields. In general, once given the matrix
\pref{g3diag} of the SC couplings, we are interested to the case where
there are two bonding eigenvalues $\L_1,\L_2$ and one antibonding one
$-\L_3$. The $T=P_{\a,\b,\varphi}R$ transformation in Eq.\ \pref{grot} is
defined through the rotation $R$ which diagonalizes $g$ and the matrix
$P_{\a,\b,\varphi}$ consisting in a 3D Poincar\'e transformation in the
$h$-space with $h_{1,2}$ being the spatial $(x,y)$ dimensions and $h_3$ the
time $(t)$ one:

\begin{widetext}
\begin{equation}
\lb{matrixu}
P_{\a,\b,\varphi}=\begin{pmatrix}\frac{1}{\sqrt{\Lambda_{1}}} & 0 & 0\\
0 & \frac{1}{\sqrt{\Lambda_{2}}} & 0\\
0 & 0 & \frac{1}{\sqrt{\Lambda_{3}}}
\end{pmatrix}
R_{\alpha}^{xy}H_{\varphi}^{\vec{v}_{xy}(\b),t}
\begin{pmatrix}\sqrt{\Lambda_{1}} & 0 & 0\\
0 & \sqrt{\Lambda_{2}} & 0\\
0 & 0 & \sqrt{\Lambda_{3}}
\end{pmatrix}
\quad\alpha,\beta \in[0,2\p]\quad\varphi\in\mathbb{R}
\end{equation}
\end{widetext}
Here $R_\alpha^{xy}$ is an ordinary rotation of angle $\a$ (with respect to
the $x$ axis) in the $(x,y)$ plane, while
$H_{\varphi}^{\vec{v}_{xy}(\b),t}$ is a hyperbolic rotation of angle
$\varphi$ in the plane identified by the $t$ direction and by the versor
$\vec{v}_{xy}(\beta)$ of the $(x,y)$  plane, $\beta$ being the angle
with respect to $x$.
After the HS decoupling the equivalent of the action
\pref{snew} will read:
\bea
\lb{s3bands}
S&=&S_0+\int d\tau d\bx\,
\frac{|h_{1}(x)|^{2}}{\Lambda_{1}}+\frac{|h_{2}(x)|^{2}}{\Lambda_{2}}+\frac{|h_{3}(x)|^{2}}{|\Lambda_{3}|}\nn\\
& &-\left(h_{1}^{^{*}}\psi_{1}+h_{2}^{^{*}}\psi_{2} +h.c.\right)-i\left(h_{3}^{^{*}}\psi_{3}+h.c.\right)
\eea
We can then proceed as in the two-band case, having in mind that now
the HS field $h_3$ associated to the antibonding channel will enter with
an imaginary unit in both the saddle-point Green's function \pref{defgbar}
and the self-energy \pref{defsigma}, playing then the role of $h_2$
for the two-band case. We observe that the $T$ matrix depends on three
parameters, i.e. the rotation angles $\a,\b,\varphi$ of the matrix \pref{matrixu}
above: they are fixed (self-consistently) by the three conditions \pref{fix3} above which
are used to decouple the saddle-point equations for the HS fields. In the ordinary SC phase, when only the $h_1$
field has a finite saddle-point value, the derivation of the action is
a straightforward extension of the calculations presented in Sec. II.

Now we prove briefly that, if exists $T_{TRSB}$ such that Eq. \pref{ttrsb} holds, then 
 the system undergoes a second-order phase transition to a TRSB
 phase. In particular, $\bar{h}_2$ will emerge purely imaginary. 
 First of all we recall that the $T$ transformation is used to put the
 saddle-point equations \pref{usual} in diagonal form, see Eq.\
 \pref{diagt}. In the ordinary TRS phase only one element of the matrix
 $T\hat \Pi T^T-\hat\L^{-1}$ vanishes, while at $T\leq T_{TRSB}$  Eq. \pref{ttrsb}
 holds and a second element vanishes. In this situation the $h_1$ and $h_2$
 spaces are degenerate and any additional rotation $\a$ in the transformation matrix
 $P_{\a,\b,\varphi}$ will leave the result unchanged. 
 Hence, if we define in general $\bar h_i=R_i+iI_i$ we can use the
 parameter $\a$, along with the U(1) gauge symmetry,
 to impose $I_1=R_2=0$. Indeed, even if at some $T<T_{TRSB}$ one has
 $I_1,R_2\neq0$ we can pass to an other solution $(\vec{R}',\vec{I}')$ with
 vanishing $I'_1$ and $R'_2$ by means of the transformation
\begin{widetext} 
\begin{equation}
\lb{gauge+rot}
\hat{|\Lambda|}^{-1/2}\begin{pmatrix}\cos\alpha & -\sin\alpha & 0 & 0\\
\sin\alpha & \cos\alpha & 0 & 0\\
0 & 0 & \cos\alpha & -\sin\alpha\\
0 & 0 & \sin\alpha & \cos\alpha
\end{pmatrix}\hat{|\Lambda|}^{1/2}
\begin{pmatrix}\cos\vartheta & 0 & -\sin\vartheta & 0\\
0 & \cos\vartheta & 0 & -\sin\vartheta\\
\sin\vartheta & 0 & \cos\vartheta & 0\\
0 & \sin\vartheta & 0 & \cos\vartheta
\end{pmatrix}\begin{pmatrix} R_{1}\\
R_{2}\\
I_{1}\\
I_{2}
\end{pmatrix}=\begin{pmatrix}R'_{1}\\
0\\
0\\
I'_{2}
\end{pmatrix}
\end{equation}
\end{widetext} 
where $\vartheta$ is the U(1) angle.  Once established the possibility
to choose $R_2=0$ we should prove that at $T<T_{TRSB}$ the ground state favors
indeed a finite value of the imaginary part of $\bar h_2$. Indeed, even if
at $T=T_{TRSB}$ Eq.\ \pref{ttrsb} is satisfied, at lower temperatures there
are still three possibilities: (i) $I_2$ remains zero and Eq. \pref{ttrsb} does
not hold anymore, so that $R_1$ reamins the only order parameter; (ii)
$\bar h_2$ opens with a real component $R_2$  only or (iii)
$\bar{h}_2$ opens with an finite imaginary component $I_2$, and a TRSB
phase is established. To show that the case (iii) is the ground state we make
use of the fact that the imaginary fluctuations of $h_2$ at $q=0$ become
massless at $T=T_{TRSB}$, as proven in Eq.\ \pref{sphase3q0} above. Indeed,
let us write down the expansion of the action at a temperature
$T\lesssim T_{TRSB}$ with respect to the mean-field action $\tilde
S_{MF}$ computed with the solution (i), i.e. $\vec R=(R_1,0,0), \vec
I=\vec 0$. By using in
the fluctuation action the $T$ matrix and the fermionic bubble evaluated at
the expansion point and the results of Sec. III we have
\bea
S&=&\tilde S_{MF} +\frac{1}{2}
\d\vec{R}^{T}({T}\hat{\L}^{11}{T}^{T}+2\hat{\L}^{-1})\d\vec{R} +\nn\\
& & + \frac{1}{2}\d\vec{I}^{T}(\hat{T}\hat{\L}^{22}\hat{T}^{T}+2\hat{\L}^{-1})\d\vec{I} + \mathcal{O}(R^3,I^3,RI^2,IR^3)\nn\\
\lb{sexp}
\eea
where $\d\vec R,\d \vec I$ are the displacements with respect to the
expansion point, and  $\d \vec I$ includes the imaginary factor $i$ of the antibonding channel,
in accordance with the definition given above Eq.\ \pref{sphase3q0}. Observe that in Eq.\ \pref{sexp}
the linear terms do not appear since  $S_{MF}^{(1)}$ is a stationary point
for the action, and the phase-amplitude couplings are absent since above
$T_{TRSB}$ we have chosen the gauge where all the gaps are real. Using the
relations \pref{l11}-\pref{l22} for the $q=0$ values of the fermionic
bubble we can then see that $\d\vec R$ fluctuations are always costly,
so that case $(ii)$ leads to an increase of the energy. On the other hand,
thanks the identity $\hat\L_{22}(q=0)=-2\hat\Pi$ the 
$\d\vec I$ fluctuations have explicitly the form:
\begin{equation}
\d\vec{I}^{T}diag(0,-a,-b)\d\vec{I} =a'(T-T_{TRSB}) I_2^2+b I_3^2,  \quad a,a',b>0
\end{equation}
where we used the fact that at $T=T_{TRSB}$ Eq.\ \pref{ttrsb} holds
and $\d I_2$ fluctuations are massless. As one can see, $\d I_3$
fluctuation increase the energy while $\d I_2$ fluctuations decrease
it, making the TRS phase unstable towards a phase with a finite $I_2$,
which leads to non-trivial phases for the gaps \pref{mfgaptrsb} and
then to a TRSB phase. This instability will be of course compensated
by higher-order terms in the expansion \pref{sexp}, that can also lead
to a finite $\d I_1$ and $\d R_2$ along with a finite $\d
I_2$. However, as we discussed above, these components can be
eliminated by the trasformation \pref{gauge+rot}, making the
definition \pref{mfgaptrsb} fully general.

We have then proven that the vanishing of a second eigenvalue of the
matrix $\hat \Pi-\hat g^{-1}$, i.e. Eq.\ \pref{ttrsb}, is a sufficient condition for a TRSB
phase, since when this happens the $\bar h_2$ field acquires a finite
imaginary part. We will now show that this is also a necessary
condition for having a TRSB phase. Let us go back to the set of
self-consistency equations \pref{usual}, and let us decompose the
matrix $\hat \Pi-\hat g^{-1}$ in its eigenvectors at a temperature
$T\leq T_c$ where the ordinary SC state is established:
\be
\lb{eqeigenv}
\hat \Pi-\hat g^{-1} =\l_2(T) {\bu}^T_2 \bu_2+\l_3(T) {\bu}^T_3 \bu_3
\ee
where $\l_i(T)$ and $\bu_i$ are real, since the
matrix $\hat \Pi-\hat g^{-1}$ is real and symmetric. Here we used the fact that below $T_c$ one eigenvalue vanishes,
allowing for a finite solution $\vec \D$. In general,  $\vec \D$ is a
vector of complex numbers, that satisfy the self-consistency equation
\pref{usual}, i.e.:
\bea
\lb{re}
& &\l_2(T) (Re \vec\Delta\cdot{\bu}_2) \bu_2+\l_3(T) (Re \vec\Delta\cdot{\bu}_3) \bu_3=0\\
\lb{im}
& &\l_2(T) ( Im\vec\Delta\cdot{\bu}_2) \bu_2+\l_3(T) ( Im\vec \Delta\cdot{\bu}_3) \bu_3=0
\eea
Since $\bu_2$ and $\bu_3$ are orthogonal, the above equations imply
that the vectors $Re\vec \D$ and $Im \vec \D$ are also orthogonal to
both $\bu_2$ and $\bu_3$. Thus, either one of the two vanishes, or
they are paralell to $\bu_1$. In all these cases the gaps have all the
same phases, and then the state is TRS. On the other hand, when one
additional eigenvalue vanishes in Eq.\ \pref{eqeigenv}, say $\l_2=0$,
then  $Re\vec \D$ and $ Im \vec\D$ belong to the two-dimensional subspace spanned by $\bu_1$ and
$\bu_2$, so that their phases can be complex. In this respect, it can
also be instructive to show how simple geometrical arguments can be
used to establish if a TRSB phase exists a $T=0$ for a generic
coupling matrix. Let us start from Eq.\ \pref{eqeigenv} with $\l_2=$,
and let us redefine $\bV=\sqrt{\l_3}\bu$, so that $\hat \Pi-\hat
g^{-1}$ is written explicitly as a projector:
\be
\lb{cond1}
\hat{\Pi}-\hat{g}^{-1} = \bV^T \bV.
\ee
This equation allows one to determine $\bV$ in terms only of the
couplings: indeed, it gives explicitly:
 \begin{equation}
\lb{ginv}
g^{-1}=-\begin{pmatrix}V_1^{2}-\Pi_{1} & V_1V_2 & V_1V_3\\
V_1V_2 & V_2^{2}-\Pi_{2} & V_2V_3\\
V_3V_1 & V_3V_2 & V_3^{2}-\Pi_{3}
\end{pmatrix}
\end{equation}
Thus, temperature or bands
parameters, which only enter via the Cooper bubbles, do not affect the
forbidden direction $\bV$, that is fully determined as (let
$\hat{G}$ be the inverse of $\hat{g}$):
\bea
\lb{v11}
V_1^{2} & = & -\frac{G_{12}G_{31}}{G_{23}},\\
V_2^{2} & = & -\frac{G_{21}G_{32}}{G_{13}},\\
V_3^{2} & = & -\frac{G_{31}G_{23}}{G_{12}}.
\eea
Observe also that $V$ can be identically zero only if all the
interband couplings vanish. Thus is any real multiband system, where at
least one interband coupling is finite, there must exist one forbidden
direction. In the two-band case this guarantees that the second
eigenvalue of the matrix $\hat \pi-\hat g^{-1}$ never vanishes. 
Once $\bV$ is known, the self-consistency equations for the gap
amplitudes are also determined by the diagonal terms in Eq.\
\pref{ginv}, that give explicitly:
\begin{equation}
\lb{pinew}
\Pi_{i}=G_{ii}+V_{i}^{2}=G_{ii}-\frac{G_{ik}G_{ji}}{G_{kj}}\qquad\left(\forall i,\, k\neq j\neq i\right).
\end{equation}
As one can see, in the TRSB phase the equations for the gap amplitudes
in the various bands decouple, and they all reduce to a single-band
BCS equation with different effective couplings. This allows us also
to express in the self-consistency equations \pref{re}-\pref{im} the
gap amplitudes as a function of the Cooper bubbles, by inverting the
relation $\Pi_i(T=0)=N_i\mathrm {asinh}(\o_0/|\D|_i)$. Thus, Eq.s
\pref{re}-\pref{im} above reduce in the TRSB state to
\bea
\lb{vre}
\vec{V}\cdot{Re\vec{\D}}=0\Ra\quad \sum_lV_l\frac{\o_0\cos\bar{\vartheta_l}}{\sinh{\Pi_l/N^l_0}}=0,\\
\lb{vim}
\vec{V}\cdot{Im\vec{\D}}=0 \Ra\quad \sum_lV_l\frac{\o_0\sin\bar{\vartheta_l}}{\sinh{\Pi_l/N^l_0}}=0.
\eea
As usual, gauge invariance allows us to fix one of the phases to
zero, say $\vartheta_3=0$. Thus, Eqs.\ \pref{vre}-\pref{vim} determine
the two remaining non-trivial phases $\vartheta_1,\vartheta_2$ in terms only
of the coupling matrix and band parameters, given by Eqs.\
\pref{v11}-\pref{pinew} above. For example, in the case of the
coupling matrix considered in Eq.\ \pref{defg3} one has for the the inverse matrix (in
units $N_l=N=1$):
\begin{equation}
G=-\frac{1}{2}\begin{pmatrix}-\frac{1}{\eta} & \frac{1}{\eta} & \frac{1}{\lambda}\\
\frac{1}{\eta} & -\frac{1}{\eta} & \frac{1}{\lambda}\\
\frac{1}{\lambda} & \frac{1}{\lambda} & -\frac{\eta}{\lambda^{2}}
\end{pmatrix},
\ee
so that one immediately determines
\bea
& &V_1^2=V_2^2=\frac{1}{2\eta}, \quad V_3^2=\frac{\eta}{2\l^2},\\
& &  \Pi_1=\Pi_2=\frac{1}{\eta}, \quad \Pi_3=\frac{\eta}{\l^2}.
\eea
Thus, Eqs.\ \pref{vre}-\pref{vim} lead to
$\vartheta_1=-\vartheta_2$ where 
\be
\cos \theta_1=-\frac{\eta}{2\lambda}\frac{\sinh \Pi_1}{\sinh
  \Pi_3}\simeq -\frac{\eta}{2\lambda} \exp\left(\frac{1}{\eta}-\frac{\eta}{\l^2}\right),
\ee
which admits, at fixed $\eta$, a  finite solution only for $\l\leq
\l_{cr}(\eta)$, see the phase diagram shown in Fig.\ \ref{fig-pd}.


\begin{thebibliography}{99}
\bibitem{footnote_T}
Notice that  the $T$ matrix is not unitary, so
the two combinations of fields $\psi_1$ and $\psi_2$ are not
orthogonal. However, this is not a problem,  since only the $\psi_1$ combination of
fermionic field will turn out to be the relevant SC channel to describe the
mean-field solution.
\end{thebibliography}
\end{document}